\documentclass[sigconf, 10pt, colorlinks]{acmart}
\setcopyright{acmlicensed}
\usepackage{balance}
\usepackage{subcaption}
\usepackage{colortbl}
\usepackage{algorithm}
\usepackage{algorithmic}
\usepackage{soul}
\usepackage{listings}
\newcommand{\tabfill}{\textsf{AutoTableComplete} } %textsc
\newcommand{\nop}[1]{}

\definecolor{Gray}{gray}{0.9}

\begin{document}

\title{Automatic Table completion using Knowledge Base}

\author{Bortik Bandyopadhyay}
\email{bandyopadhyay.14@osu.edu}
\affiliation{%
  \institution{The Ohio State University}
  \city{Columbus}
  \state{Ohio}
}

\author{Xiang Deng}
\email{deng.595@osu.edu}
\affiliation{%
  \institution{The Ohio State University}
  \city{Columbus}
  \state{Ohio}
}

\author{Goonmeet Bajaj}
\email{bajaj.32@osu.edu}
\affiliation{%
  \institution{The Ohio State University}
  \city{Columbus}
  \state{Ohio}
}

\author{Huan Sun}
\email{sun.397@osu.edu}
\affiliation{%
  \institution{The Ohio State University}
  \city{Columbus}
  \state{Ohio}
}

\author{Srinivasan Parthasarathy}
\email{srini@cse.ohio-state.edu}
\affiliation{%
  \institution{The Ohio State University}
  \city{Columbus}
  \state{Ohio}
}

\begin{abstract}
Table is a popular data format to organize and present relational information. Users often have to manually compose tables when gathering their desiderate information (e.g., entities and their attributes) for decision making.
In this work, we propose to resolve a new type of heterogeneous query viz: \textit{tabular query}, which contains {a natural language query description}, {column names of the desired table}, and {an example row}.
We aim to acquire more entity tuples (rows) and automatically fill the table specified by the \textit{tabular query}.
We design a novel framework \tabfill which aims to integrate schema specific structural information with the natural language contextual information provided by the user, to complete tables automatically, using a heterogeneous knowledge base (KB) as the main information source.
Given a tabular query as input, our framework first constructs a set of candidate chains that connect the given example entities in KB.
We \textit{learn to select the best matching chain} from these candidates using the \textit{semantic context from tabular query}.
The selected chain is then converted into a SPARQL query, executed against KB to gather a set of candidate rows, that are then ranked in order of their relevance to the \textit{tabular query}, to complete the desired table.
We construct a new dataset based on tables in Wikipedia pages and Freebase, using which we perform a wide range of experiments to demonstrate the effectiveness of \tabfill as well as  present a detailed error analysis of our method.\looseness=-1
\end{abstract}


\begin{comment}
\begin{CCSXML}
<ccs2012>
<concept>
<concept_id>10002951.10003317.10003347.10003352</concept_id>
<concept_desc>Information systems~Information extraction</concept_desc>
<concept_significance>500</concept_significance>
</concept>
<concept>
<concept_id>10002951.10003317.10003371.10010852</concept_id>
<concept_desc>Information systems~Environment-specific retrieval</concept_desc>
<concept_significance>500</concept_significance>
</concept>
</ccs2012>
\end{CCSXML}

\ccsdesc[500]{Information systems~Information extraction}
%\ccsdesc[500]{Information systems~ Environment- specific retrieval}

\keywords{Table completion, Machine Learning, Knowledge Base}
\end{comment}

\maketitle

\section{Introduction}
\label{section:introduction}

Users may issue queries about entities related to specific topics, as well as their attributes or multi-ary relationships among other entities~\cite{Pimplikar2012ATQ}, to gather information and make decisions.
Tabular representations are a natural mechanism to organize and present latent relational information among a set of entities~\cite{Vessey1991}.
Tables are widely used in existing commercial products (e.g.: Microsoft Excel PowerQuery\footnote{\url{http://office.microsoft.com/en-us/excel/download-microsoft-power-query-for-excel-FX104018616.aspx}}) and within novel frameworks (e.g.:~\cite{Pimplikar2012ATQ,Yakout2012IEA,yang2014finding,ZhangSmartTable}).
Recently, search engines (like Google) also tend to return tables as answer to list-seeking user queries~\cite{balakrishnanapplying, Huang2019CFR} thereby making tabular representation even more popular.\looseness=-1

Oftentimes the user knows precisely the kind of information they are looking for (e.g.: topic entity) and can even guide the information gathering process with a correct example, due to their (partial) familiarity with the topic entity, coupled with the practical need to aggregate additional relevant information about it with least manual effort.
Imagine that a prospective freshman wants to know the names of some distinguished television personalities, who are alumni of colleges that are located in a specific part of a country (e.g. west coast), to better understand the college's environment for students interested in drama.
The parent maybe curious to know the list of airlines and their corresponding destination airports that fly from the airport closest to such a college campus.
In both cases, it is plausible that the user already has some seed information of the desired table which can be provided as guidance.
In turn, the user expects the table to be auto-populated by an application/service, which can accept the precise input as well as utilize it effectively for this task.\looseness=-1

\begin{figure}[!htb]
\includegraphics[width=\linewidth]{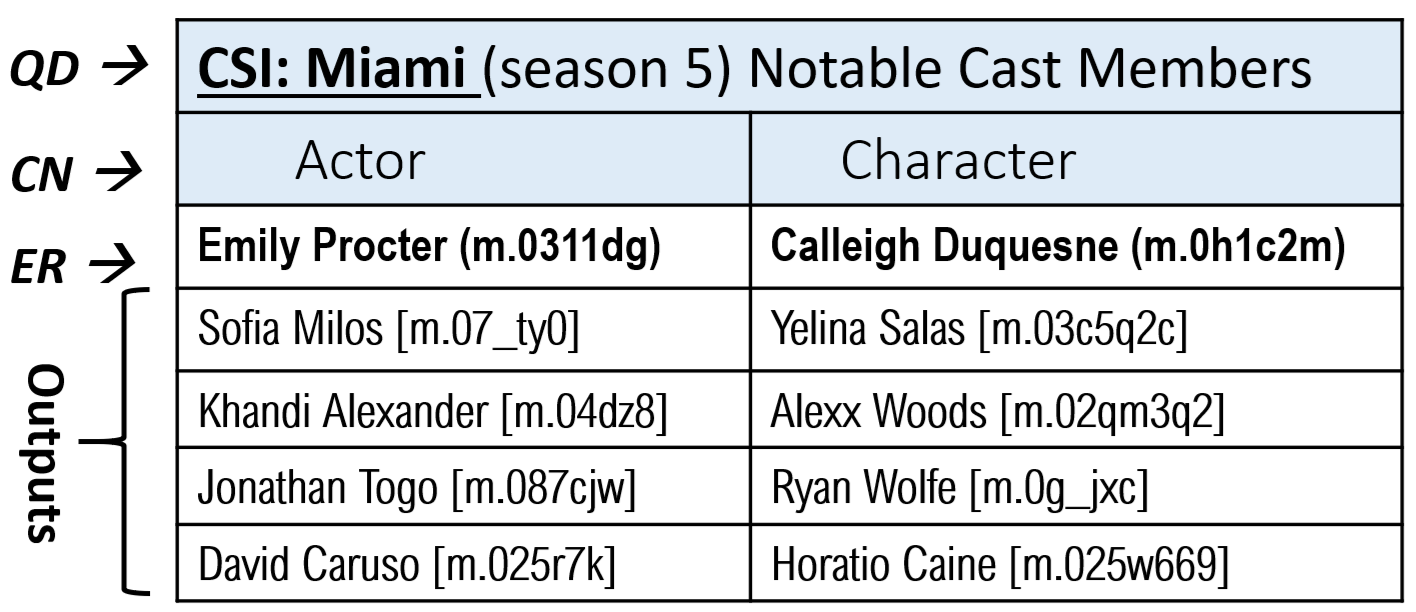}
\caption{An example \textit{tabular query} $<$QD, CN, ER$>$. The user is interested to know about ``\textit{Season 5 Notable Cast Members}'' of the Subject Entity ($SE$) = ``\textit{CSI: Miami}''. The specific attributes, i.e., a subset of information the user is looking for, are expressed by the two columns: ``\textit{Actor}'' and ``\textit{Character}''. ER is provided as an example row to further illustrate the intent. We aim to augment additional rows that best meet the user's information requirement.\looseness=-1}
\label{fig:tabularq}
\end{figure}

We empower the user to express their specific information needs through a new type of query called \textit{tabular query}.
Such a query contains {a natural language query description (QD), column names (CN) to specify interesting attributes, and one example row (ER)}. 
Figure~\ref{fig:tabularq} shows an example of a tabular query.
While there are many possible relationships associated with the topic entity ``\textit{CSI: Miami}'', the specific information the user may be interested in could be the $<$Actor - Character$>$ relationship pair, expressed by CN.
Tabular queries can be realized using custom plug-ins for office utility products (e.g.: Microsoft Excel) or through a custom web portal.
In this paper, we study \textit{tabular query resolution} which aims to address a user-provided tabular query by automatically filling the desired table with relevant output rows.\looseness=-1

An intuitive approach is to use the natural language part of \textit{tabular query} i.e., QD as input to search engines (like Google) or state-of-the-art \textit{table search/augmentation/generation frameworks}~\cite{cafarella2008webtables,nguyen2015result,ZhangAdhocTable,venetis2011recovering,Zhang2018OTG}.
However, the problem with such approaches is that the user has no control over the schema of the output table, as expressed using $<$CN, ER$>$ in a \textit{tabular query}.
Thus, either extra information is retrieved possibly overwhelming the user with additional unwanted columns, or some information is missed by not gathering columns of interest.
Recent works that try to automatically generate/predict the relevant columns~\cite{Zhang2018OTG,Huang2019CFR} of the output table, might be well suited for a rather exploratory information gathering task.
However, tabular queries enable the user, who has a well-defined information requirement, to specify {the general topic as well as the structure} of the desired table, thereby allowing more granular control on the output, when compared to traditional natural language queries.
Such control over the schema of the output table can significantly reduce post-processing burden from the user, who can now focus on leveraging the completed table for the intended task, rather than editing the columns of the generated table.\looseness=-1

A \textit{tabular query} contains both contextual and structural information of the desired table. While the existing techniques are built to lever the contextual information, the latent structural relationship between columns is not modeled, as it is very difficult to represent such a \textit{latent relation}. Jointly leveraging the content and structure information of the input tabular query for completion of a user specified multi-row multi-column table is a challenging task that we address.\looseness=-1

To this end, we propose a novel framework named \tabfill that levers both the content and structural information provided by the user to automatically complete rows of the table.
\tabfill has three main components: \textit{Query Preprocessor} (QP), \textit{Query Template Selector} (QTS) and  \textit{Candidate Tuple Ranker} (CTR).
QP is a light-weight component with the key task of converting the user provided \textit{tabular query} into a relevant format that can be utilized by the QTS and CTR modules.
Given an example row (ER) in the processed tabular query, QP composes a set of candidate relational chains, which are essentially multi-hop labeled meta-paths, connecting entities of the example row in a Knowledge Base (KB).
We develop a plug-and-play learning module as part of QTS that learns to automatically select the best candidate chain using the \textit{tabular query}.
The best-selected chain is translated into a SPARQL query, and executed against the KB to gather a set of candidate entity tuples as additional rows to complete the table. The retrieved entity tuples are ranked by a learning-based CTR module in order of their relevance to the {tabular query}, using additional information from the KB.
To summarize, our contributions are:\looseness=-1
\begin{itemize}
    \item To the best of our knowledge, \tabfill is the first framework that jointly levers structural information between entities in table columns with unstructured natural language query context, to automatically complete tables for the user-provided tabular query.
    \item We represent the latent information between columns using \textit{KB relational chains}, and use machine learning based components for learning-to-complete tables.
    \item We construct a novel dataset from the Wiki Table corpus~\cite{bhagavatula2015tabel} to systematically evaluate our framework.
    \item We conduct a wide range of experiments to demonstrate the effectiveness of our framework, followed by a detailed error analysis to identify future focus areas.
\end{itemize}

To facilitate future research, the resources will be made available at : {github.com/bortikb/AutoTableComplete}
\section{Problem Statement}
\label{section:kbtablefill}

We study the problem of automatically completing a partially filled table provided as input through a \textit{tabular query} (defined below).
Specifically, we focus on tables which are 1) composed of only entities that can be mapped to a knowledge base (KB), and 2) the leftmost column consists of unique entities i.e., it is the primary key column.\looseness=-1

\textbf{\noindent Tabular Query:} We define a tabular query to have three parts:  1) \textit{the natural language Query Description} (QD) that contains the Subject Entity (SE) of the table, along with the context of user's information requirement (QIS) about the subject entity 2) \textit{Column Names} (CN) for $l$ columns expressed in natural language and 3) one \textit{Example Row} (ER), where $(ER_1, ER_2, .., ER_l)$ is a $l$ column tuple, such that $ER_1$ is the column 1 value, $ER_2$ is the column 2 value etc..\looseness=-1

We assume that all parts of the tabular query are non-empty and correct. Thus the natural language sub-parts $<$QD,CN$>$  jointly conveys the same information as ER.\looseness=-1

\textbf{\noindent Tabular Query Resolution:} Given a tabular query $<$QD, CN, ER$>$ and a KB as the main information source, our task is to complete the table with KB entities as cell values, such that the final table best represents the user's information need.\looseness=-1

As pointed out by~\cite{Huang2019CFR}, most of the real world table synthesis or compression applications often have the number of relevant facts about entities (columns of the output table) set to 3, so that the rendered table can fit on the phone screen.
This observation makes us believe that an effective solution for a smaller subset of columns can help generalize our approach to the majority of the daily use cases.
Thus, without loss of generality, we focus on completing tables with exactly 2 columns (i.e., $l$ = 2), where the leftmost column is the primary key.
However, our method can be easily extended to tables with $l$ columns (where $l > 2$), by first constructing $(l-1)$ 2-column tables with the primary key as the left-most column for each such tables using our strategy, and then joining all these 2-column tables using the primary key column, to generate the final table. 
Thus, in this work we mainly focus on jointly leveraging the content (provided by QD, CN \& ER) and structure (provided by CN \& ER) information of \textit{tabular query} for the table completion task, which necessitates the use of a knowledge base (KB) that we discuss next.\looseness=-1

\section{Tabular Query Resolution}

\begin{figure*}[!htb]
\includegraphics[width=\textwidth]{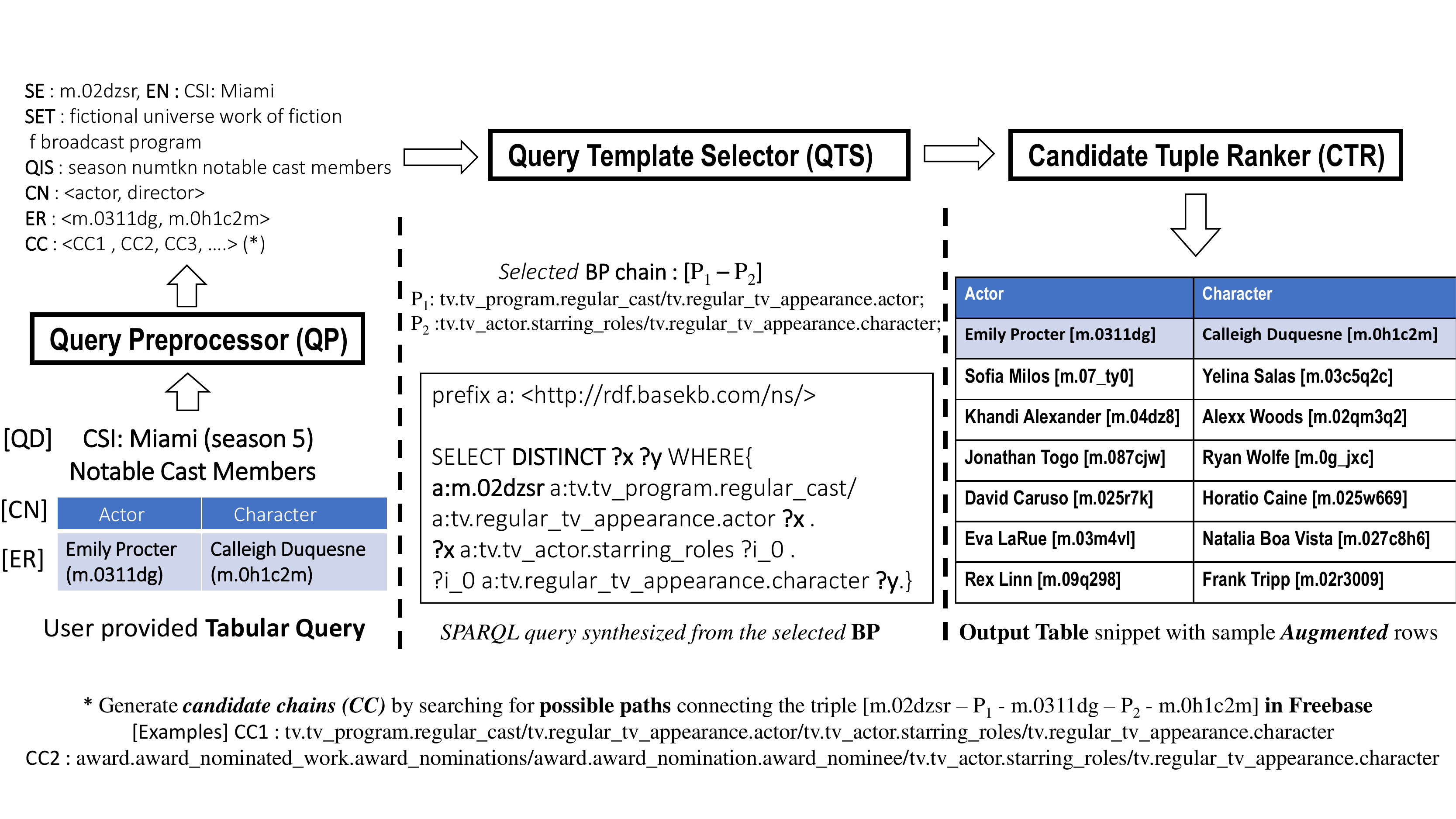}
\vspace{-30pt}
\caption{The user provides $<$QD, CN, ER$>$  as \textit{tabular query}. Besides performing standard text processing tasks, {QP} module extracts and links entities to KB. It uses these linked KB entities to generate a set of candidate chains (CC) that connect those entities in the KB via meta-paths. {QTS} uses the pre-processed tabular query to select the best path (BP) from CC. BP is used to construct a SPARQL query that is executed against KB to gather a set of candidate tuples $(?x , ?y)$. These candidates are ranked by the {CTR} using an ensemble of features. The final output table of our framework is a ranked list of candidate rows.\looseness=-1}
\vspace{-0.25cm}
\label{fig:e2eframework}
\end{figure*}

\subsection{Analyzing Tabular Query}
We start by analyzing various types of information available to us from the different sub-parts of \textit{tabular query}.
The first component is the \textit{natural language query description}, containing the \textit{subject entity} of the table.
The remaining query description contains natural language contextual information.
Both column names and example row contains relationship information between column values of the desired table. 
In column names such as ``Actor'' - ``Character'', the relationship information (i.e., \textit{characters played by actors}) is conveyed by the natural language tokens, while in comparison, the relationship between $ER_1$ and $ER_2$ in the example row is latent and relatively harder to represent and use.
We primarily focus on using a knowledge base (KB) to represent the latent semantic relationships between entities.\looseness=-1

\subsection{Using KB to Resolve Tabular Queries}
A knowledge base considered in this work is a set of assertions $(e_1, p, e_2)$, where $e_1$ and $e_2$ are respectively subject and object entities such as \textsf{Emily Procter} and \textsf{Calleigh Duquesne}, and $p$ is the binary predicate between them such as \textsf{tv.tv\_actor. starring\_roles / tv.regular\_tv\_appearance.character}. A knowledge base is often referred to as a \textit{knowledge graph} where each node is an entity and each edge is directed from the subject to the object entity and labeled by the predicate. Similar to previous work \cite{Sun2016TCS}, we define a meta-path between two entities as a sequence of predicates on edges that connect the two entities in KB. Even though the set of predicates for a KB is pre-defined (and hence static), they can be combined in various order to form valid meta-paths, to express a rich and diverse set of semantic relationships between any pair of connected entities in KB.\looseness=-1

In our framework, we use meta-paths to represent the latent relationships between example entities $ER_1$ and $ER_2$. 
To decide the best meta-path for each tabular query, the predicate names on each meta-path can be compared with other information given in the tabular query by a machine learning component.
Once the best meta-path is decided, we can convert it into a SPARQL query and execute it against the KB to obtain more pairs of entities that hold the relationship. The returned entity pairs will be effectively utilized to fill the table as answer to the tabular query.
Any of the publicly available knowledge bases (e.g.: Freebase~\cite{bollacker2008freebase}, DBPedia~\cite{auer2007dbpedia}, Wikidata~\cite{Vrandecic2014WFC} etc.), given their own characteristics~\cite{farber2015comparative}, can be used as a KB in our framework.
In this work, we choose Freebase for meta-path based relation representation between entity pairs, as it is very popular for its richness, diversity, and scale in terms of both structure and content ~\cite{farber2015comparative}, and has been widely employed as the information source to answer natural language questions~\cite{yih2015semantic,dong2015question,diefenbach2018core}.

%Unlike some KB like DBPedia, which contains only binary relations, Freebase contains both binary and non-binary relations~\cite{diefenbach2018core}, thereby allowing more complex relationships between entities (e.g. temporal relation) to be expressed.
%11230937_1

\subsection{Framework Overview}
\label{section:framework}

Our proposed framework \tabfill is shown in Figure~\ref{fig:e2eframework}. We use the user provided tabular query shown in Figure~\ref{fig:tabularq} to illustrate the inputs and outputs of each component. To summarize, a {tabular query} is first pre-processed to extract the relevant inputs, which are then sent through individual modules to produce a ranked list of rows as the output table. Given a tabular query, we pre-process it as follows:
\begin{enumerate}
    \item \textbf{Query Description (QD):} This is a natural language text description of the user's query intent. We process QD to generate the following 3 inputs viz: \textbf{Subject Entity, Subject Entity Types, and Query Intent String}. A table is typically centered around a topic of interest, which in this case, is present in QD as Subject Entity (SE). We extract $SE$ using a state-of-the-art entity linking tool, which maps it to a unique Freebase machine id (i.e., mid) = \textit{``m.02dzsr"} with \textit{``CSI: Miami"} as Entity Name (EN). The remaining text after removing EN from QD is the Query Intent String (QIS) which in this case is \textit{``season numtkn notable cast members"}, where ``numtkn" is a unified token for all numbers. The Freebase entity type\footnote{We select the most specific type when there are many.} of \textit{``m.02dzsr"} (\textit{``fictional\_universe.work\_of\_fiction"}) is processed and concatenated with its fine-grained type (\textit{``f.broadcast\_ program"}) from~\cite{ling2012fine}, to create Subject Entity Types (SET).\looseness=-1% where ``f'' is used to distinguish the fine grained type from the Freebase type.
    \item \textbf{Column Names (CN):} Column names are generally natural language tokens or phrases. For example, \textit{``Actor'' -  ``Character''} are provided as CN in Figure~\ref{fig:e2eframework}, {implying that the user wants to find actors and the corresponding characters played by them, given the SE and QIS.}  Note that natural language allows for potentially many different ways to convey the same meaning of the query description and column names. For example, users may also use \textit{``Actor'' - ``Role''} or \textit{``Name'' - ``Role''} as column names to represent the same relationship as \textit{``Actor'' - ``Character''}. Therefore, interpreting the real semantic meaning between the desired columns is one of the challenges addressed by our framework. %\vspace{-0.15cm}
    \item \textbf{Example Row (ER):} We assume the user can provide one correct example row $(ER_1, ER_2)$ of the output table, e.g.,  \textit{(Emily Procter, Calleigh Duquesne)} in Figure~\ref{fig:tabularq}. The given row is linked to Freebase entities, e.g., entities with mid's \textit{(m.0311dg, m.0h1c2m)} in the running example. We extract meta-paths connecting the triple ($SE$, $ER_1$, $ER_2$) in Freebase, which form the candidate chain set $CC$.~\footnote{We use path and chain interchangeably.} A subset of $CC$ is shown in Figure~\ref{fig:e2eframework}. The meta-paths connecting the triple, denoted as $[P_1 - P_2]$, are obtained by concatenating $P_1$ and $P_2$, where $P_1$ is a meta-path between ($SE$, $ER_1$) and $P_2$ is a meta-path between ($ER_1$, $ER_2$). Each meta-path is a sequence of predicate names on edges connecting the corresponding entity pairs. The predicate names are treated as the textual representation of the latent structural relationships between entities constituting the table.%\vspace{-0.15cm}
\end{enumerate}

Thus the user provided input $<$QD, CN, ER$>$ is pre-processed to generate $<$QIS, CN, SET, CC$>$, which will be used by the {Query Template Selector} component to select the best path (BP), i.e., the one that best matches with the user provided information in $<$QIS, CN, SET$>$.
In Figure~\ref{fig:e2eframework}, the BP is highlighted as an output of the QTS and is easily transformed into a SPARQL query  which is then executed against Freebase to retrieve a set of candidate entity tuples.
The {Candidate Tuple Ranker} component ranks the retrieved candidate tuples based on a set of structural and semantic features computed using the information from $<$QIS, SE, SET, CN, BP, ER$>$, Freebase, and an external Wikipedia text corpus\footnote{https://en.wikipedia.org/}. To train and evaluate \tabfill, we use the WebTable corpus \cite{bhagavatula2015tabel}, consisting of millions of tables in Wikipedia pages, to construct a new data set (refer to Section~\ref{section:data}). \looseness=-1
\section{Methodology}
\label{section:methodology}
We now describe the building blocks of \tabfill in details.
%\vspace{-0.85cm}

\subsection{Query Pre-processor (QP)} 
\label{subsection:input}

The $<$QD, CN, ER$>$ is processed to extract the Subject Entity (SE) from the query description and generate the Query Intent String (QIS) (as discussed in Section~\ref{section:framework}).
The SE and example row ER $= (ER_1, ER_2)$ are mapped to Freebase entities.
This step helps us to extract the Subject Entity Types (SET) and also construct a set of candidate meta-paths CC, which we briefly describe next.

\textbf{\noindent Candidate Chain Generation:} Given SE and the example row $(ER_1, ER_2)$, we extract simple paths (no cycle) between $(SE, ER_1)$ as $P_1 = R_1^1/R_2^1/../R_m^1$ and between $(ER_1, ER_2)$ as $P_2 = R_1^2/R_2^2/../R_n^2$ in KB, where $m$ and $n$ are the lengths of the individual paths and $R_i$ is the relationship label text.
For high Recall, yet diversified, path set construction between the individual entity pairs, we adopt depth first search from source to reach the target entity along with pruning strategies, upper bounding the path length to $L$ (L = 3), i.e., ($m<=3$ and $n<=3$).
%We also apply following constraints to avoid explosion of search space: 1) Stop expanding for nodes with over 500 neighbors, this greatly reduces the search space but has minor impact on results. 2) Pruning noisy relations that are unlikely to be part of right chain.
After generating a set of $M$ simple paths between $(SE, ER_1)$ and $N$ simple paths between $(ER_1, ER_2)$, the candidate path chain set $CC$ of size $(M * N)$ is generated by simply joining $P_1$ and $P_2$, i.e., a pair of paths respectively for $(SE, ER_1)$ and $(ER_1, ER_2)$. %i.e., $P_1$ and $P_2$ respectively.
We defer a detailed discussion of the candidate chain construction step to Section~\ref{subsubsection:pathfinding}.\looseness=-1
%as part of data collection

\subsection{Query Template Selector (QTS)}
\label{subsection:qts}

\textbf{\noindent Model Input:} After pre-processing the tabular query, the input to our framework essentially consists of QIS, CN, SET and a set of candidate chains ($CC$).
The main task of the Query Template Selector module is to use the natural language context provided by QIS, SET and CN to select the best matching path (BP) from a set of candidate chains ($CC$).
To this end, we propose a learning-based sub-module, which uses $<$QIS, CN, SET$>$ as query context to compute a matching score for each candidate chain in $CC$ and select the best chain BP with the highest matching score, such that it best captures the latent relationship between the columns of the output table w.r.t. the user provided query context. There are many choices for this learning sub-module, of which we describe a Deep Neural Networks (DNN) based model here, deferring the description of other learning models to Section~\ref{subsubsection:qtslearning}. The selected BP is translated into a SPARQL query  to retrieve candidate result rows, which will form the input to the subsequent module.
\looseness=-1

\begin{figure}[!htb]
\includegraphics[width=\linewidth]{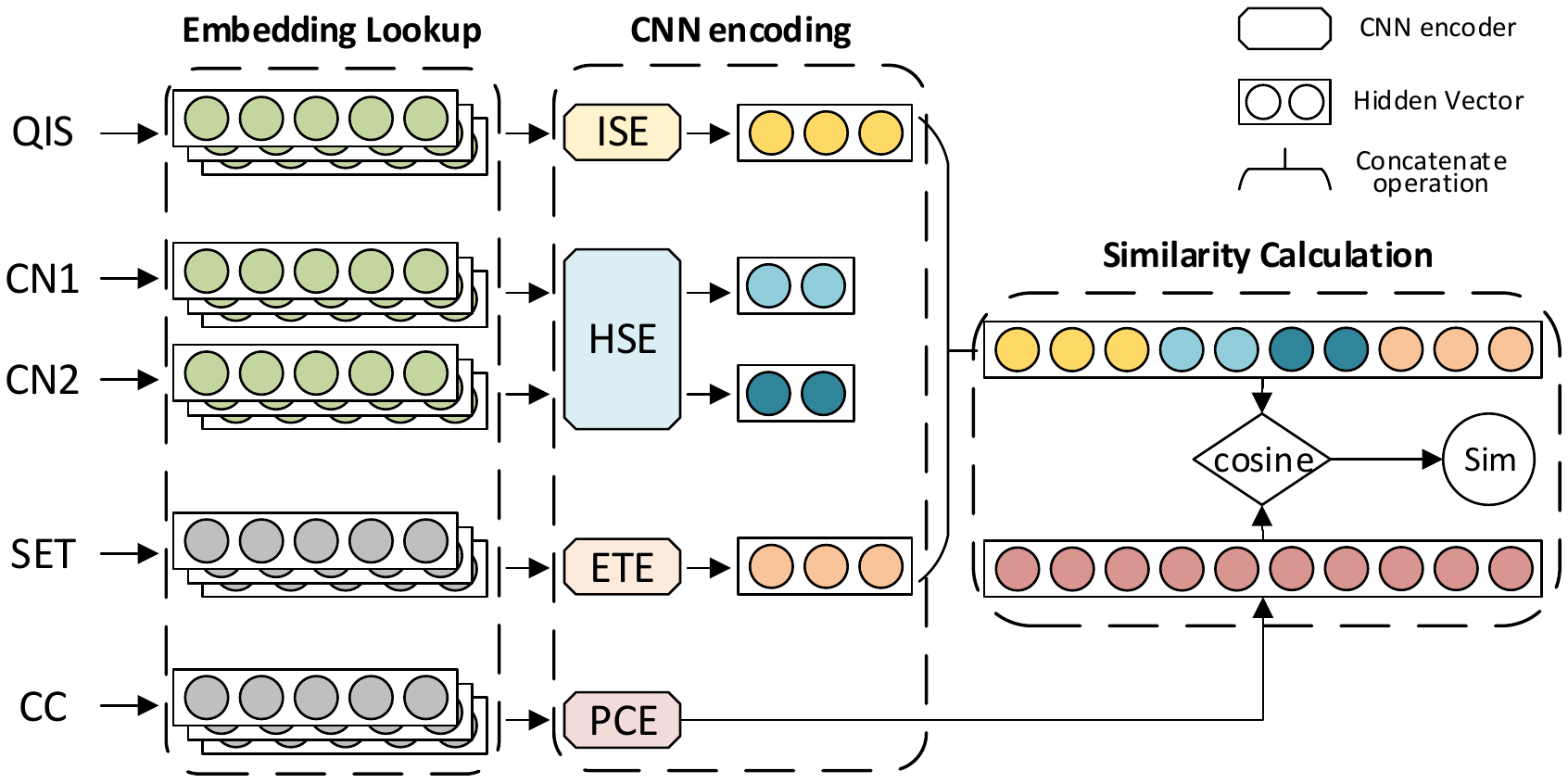}
\vspace{-20pt}
\caption{Overview of the DNN model described in Section~\ref{subsection:qts}. There are 4 CNN Encoders, one each for QIS, CN, SET, CC. The output of the ISE, HSE and ETE encoders are concatenated, ensuring that the dimension matches with the output of PCE. The cosine similarity of these output vectors is the similarity score between the tabular query and path.\looseness=-1}
\vspace{-0.25cm}
\label{fig:dnnmodel}
\end{figure}

\textbf{\noindent Model Overview:}
The architecture of the DNN model we choose is presented in Figure~\ref{fig:dnnmodel} and is similar to Siamese Networks~\cite{Koch2015SiameseNN} that have been widely used to compare pairwise text similarities.
We first encode the natural language inputs (represented by a sequence of tokens) using 4 separate Convolutional Neural Network (CNN) encoders for text data~\cite{Kim14cnn} viz: a) \textit{Intent String Encoder (ISE)}: to generate a $d_1$-dimensional vector representation of QIS; b) \textit{Header String Encoder (HSE)}: to generate 2 separate $d_2$-dimensional vector representation of CN - one for each of the individual column names ($CN_1$ and $CN_2$); c) \textit{Entity Type Encoder (ETE)}: to generate a $d_3$-dimensional vector representation of the SET; c) \textit{Path Chain Encoder (PCE)}: to generate a $d_4$-dimensional vector representation for each path in the candidate chains ($CC$).
We concatenate the three encoded vectors from ISE, HSE and ETE, and use this $(d_1 + 2 * d_2 + d_3)$-dimensional vector as the query context vector $(q)$, along with the $d_4$-dimensional encoded chain vector ($c$) for the score computation.\looseness=-1

\textbf{\noindent Model Training:} We use a positive-negative example driven objective function with hinge loss (similar to~\cite{dong2015question}) to train the DNN model.
For each table in the training set, we pick one positive chain $(p)$ and exactly $(k-1)$ negative chains $(n)$ from $CC$, assuming that each of those chains has been labeled (more details in Section~\ref{section:data}).
We generate the $d_4$-dimensional embedding of each of these $k$ chains using PCE, and then use the concatenated query context vector $(q)$ to compute $k$ pairwise cosine similarity.
$cos(q,p)$ is the Cosine similarity between $q$ and positive chain encoded vector $p$, while $cos(q, n_i)$ is the cosine similarity between $q$ and the $i$-th negative chain ($n_i$) for the current table.
These scores are used to compute hinge loss between the positive chain $(p)$ and $i$-th negative chain ($n_i$) as follows:
%$$ l(q, p, n_i) = (m - Cos(q,p) + Cos(q, n_i))_{+}$$
$$ l(q, p, n_i) = \text{max}\{0, \;\delta - cos(q,p) + cos(q, n_i)\}$$
where $\delta$ is the input margin value~\cite{dong2015question} that tries to maintain a gap of $\delta$ between the similarity scores of the positive and negative chain.
The final objective function is computed over each table ($t$) in the entire training set ($T$) as :
\begin{equation}
  \text{min}_{W} \sum_{t \in T} \sum_{p \in CC} \sum_{i = 1}^{k - 1} l(q, p, n_{i}) + \lambda * ||W||_{2}
\end{equation}
where the last term ($||W||_{2}$) is the L2 regularizer term of all trainable parameters, being weighted by a hyper-parameter $\lambda$.
The weights are learned using the back-propagation algorithm with Adam Optimizer~\cite{Kingma2015AdamAM}.

\textbf{\noindent SPARQL Query Synthesis Using Predicted Best Path:}
We use the predicted best path $[P_1 - P_2]$ in conjunction with the Subject Entity (SE) to construct a SPARQL query to select \textit{distinct} pairs of candidate tuples $(x,y)$ that satisfy the query. 
For path $P_1 = R_1^1/R_2^1/../R_m^1$ and $P_2 = R_1^2/R_2^2/../R_n^2$, where $m$ and $n$ are the number of edges in those paths and $R_i$ represents the KB predicate name of the $i$-th edge, the SPARQL query is generated as:

\begin{lstlisting}[basicstyle=\ttfamily,language=SPARQL]
prefix a: <http://rdf.basekb.com/ns/>
SELECT DISTINCT ?x ?y WHERE{
a:SE a:R$_1^1$/a:R$_2^1$/../a:R$_m^1$ ?x .
?x a:R$_1^2$/a:R$_2^2$/../a:R$_n^2$ ?y.
}   
\end{lstlisting}

Note that the set of candidate tuples $(?x, ?y)$ retrieved by executing the above SPARQL query might be very large in some cases since we do not apply any constraints from QD.
This can negatively impact the \textit{precision} of our completed tables.
To alleviate this problem, we design a Candidate Tuple Ranker module, with a diverse set of features, so that it can rank the most relevant tuples higher in the final table.\looseness=-1
\vspace{-0.25cm}

\subsection{Candidate Tuple Ranker (CTR)}
\label{subsection:ctr}

The input to this module is all the user provided pre-processed query information and a set of candidate result tuples, which are retrieved by the synthesized SPARQL query based on the best predicted path above. %by the previous module.
The task of this module is to rank the set of candidate tuples by leveraging the QIS, CN, ER, and predicted chain $[P_1 - P_2]$ so that the final ranked list best represents the user information need.
To this end, we propose a set of novel features (total 27) and use them to train a state-of-the-art learning-to-rank algorithm LambdaMART~\cite{burges2010ranknet} with the implementation from XGBoost~\cite{chen2016xgboost}, which learns an ensemble of regression tree and has shown impressive qualitative results~\cite{burges2011learning}.\looseness=-1

The \textit{first feature} tracks the number of times a particular C1 entity appears in the candidate set.
Rest of the features can be broadly divided into 3 categories viz: \textit{contextual}, \textit{semantic} and \textit{hybrid} information based.
Given the user input row $(ER_1, ER_2)$, for the $i$-th candidate tuple $(T_1^i,T_2^i)$, these 3 categories of features as extracted as follows:\looseness=-1

\textbf{\noindent Contextual information based features (Total 8)} :  For each entity, we extract the \textit{common.topic.text\_description} from Freebase as the main contextual information, which is used to compute Jaccard similarity of the BOW representation as well as Cosine similarity using pre-trained Wikipedia word embedding, thereby yielding 2 features per technique.\vspace{-0.15cm}
\begin{enumerate}
    \item \textit{Pairwise Entity Description Match Score (4 features)}: It is computed for the pairs $(ER_1, T_1^i)$ and $(ER_2, T_2^i)$. This score captures the intra-column topic similarity based on the Freebase text description of the entities.
    \item \textit{Query Intent and Entity Description Match Score (4 features)}: It is computed for the pairs $(QIS, T_1^i)$ and $(QIS, T_2^i)$. This score captures the relevance of each column item with the user's Query Intent String (QIS).
\end{enumerate}
\textbf{\noindent Semantic information based features (Total 14)}: In Freebase, each entity has an important field called \textit{common.topic. notable\_types} and another called \textit{rdf:type} (or simply \textit{type}).
Similarly, for each relationship we can extract a source expected type ($SRC\_Type$) using the relationship prefix, as well as a target expected type ($TGT\_Type$) by looking in a field called \textit{common.topic.expected\_types}.
We compute Jaccard similarity between the BOW representation of the respective text, and wherever relevant, their corresponding Cosine similarity using pre-trained Wikipedia word embedding~\cite{Mikolov2013DRW}, using the following matching techniques.
\begin{enumerate}
    \item \textit{Pairwise Entity Notable Type Match Score (4 features)}: It is computed for the pairs $(ER_1, T_1^i)$ and $(ER_2, T_2^i)$ and these scores capture the intra-column entity notable type similarity.
    \item \textit{Pairwise Entity Type Match Score (4 features)}: It is computed for the pairs $(ER_1, T_1^i)$ and $(ER_2, T_2^i)$. These scores capture intra-column entity rdf type similarity.
    \item \textit{Entity Type and Connecting Chain Type Match Score (6 features)}: We compute similarity score between the BOW representation of the entity's notable type and the connecting chain's expected type, and utilize these scores to compute 3 features as follows:\newline
    i) $S(ER_1, TGT\_Type\_P_1) - S(T_1^i, TGT\_Type\_P_1)$ \newline
    ii) $S(ER_1, SRC\_Type\_P_2) - S(T_1^i, SRC\_Type\_P_2) $ \newline
    iii) $S(ER_2, TGT\_Type\_P_2) - S(T_2^i, TGT\_Type\_P_2) $ \newline
    Each feature is a difference of two similarity scores, one obtained by using user provided input example and other obtained from the candidate tuple. Each score is computed between the entity of a particular column and the corresponding expected type (either $SRC\_Type$ or $TGT\_Type$) of the QTS predicted $[P_1 - P_2]$ relationship connecting with that entity, depending on which column and connecting chain is used.
\end{enumerate}
\textbf{\noindent Hybrid information based features (Total 4):} \textit{Pairwise Column Name and Type Match Score} features are hybrid due to their source of information.
These features capture the difference in similarity score of the natural language column names ($C1\_Name$, $C2\_Name$) and corresponding Freebase entity's notable type of that column, between the user example and the current candidate tuple. We compute: \newline
    1) $S(ER_1\_Type,C1\_Name)$ - $S(T_1^i\_Type,C1\_Name)$ \newline
    2) $S(ER_2\_Type,C2\_Name)$ - $S(T_2^i\_Type,C2\_Name)$. 
We use the pairwise difference between the similarity scores for respective columns as input features. For each similarity score computation, we use Jaccard similarity on the BOW representation, and Cosine similarity using pre-trained Wikipedia word embedding, to generate two separate set of features.
\looseness=-1
\section{Dataset Construction}
\label{section:data}

To train the learning based components in \tabfill and systematically evaluate them, we construct a dataset based on a large-scale table corpus and Freebase. 
Note that we do not have access to search engine query logs, which prevents us from knowing real world queries the users may be interested in.
Hence, for our data construction, we rely on the very popular WikiTable corpus~\cite{bhagavatula2015tabel}, which is created and curated by humans for their information need and has been extensively used in the community to demonstrate research prototypes.
Our final data contains about 4K tables, each with 2 columns and atleast 3 rows, similar to constraints used in previous works like~\cite{bao2018table}.
We leave additional data cleaning through manual curation and further scaling up of the data for future work and currently focus on describing our data extraction heuristic.\looseness=-1

\subsection{Data Source}
\label{subsection:datasource}

We start with the popular WikiTable corpus~\cite{bhagavatula2015tabel}, originally containing around 1.65M tables with differing number of columns and extracted from Wikipedia pages. 
The corpus covers a \textit{wide variety of topics} ranging from important events (e.g., Olympics) to artistic works (e.g., movies directed by a filmmaker).
The topic, as well as the content of a table, is often described by various information in the corresponding Wikipedia page and table.
As pointed out by \cite{zhang2017effective}, for each unique table, the corresponding Wikipedia Page Title and the Table Caption together can be considered as a good proxy for the user's \textbf{Query Description (QD)}.
The \textbf{Column Names (CN)} are table headers. %i.e, partially define the schema of the desired table. 
The table body consisting of $R$ distinct rows form \textbf{Result Rows (RR)}.
Since we construct our dataset from this corpus, the extracted dataset is both diverse and realistic, as the web tables were originally created and curated by humans for their information need.\looseness=-1
%, which best represents the desired table for the above definition of QD and CN.\looseness=-1
%Note that there is no explicit mention of the column set that constitutes the primary key, nor is there any clear definition of the data type and inter-relationship of the columns, which makes the schema of these tables partially defined.
%We use Freebase\footnote{https://developers.google.com/freebase/} as the knowledge base in this work.
%It contains over 43 million entities and 2.4 billion facts about them. %, which is much larger than other KBs like Yago \cite{Suchanek2008YLO} and DBpedia \cite{auer2007dbpedia}.
%An advantage of using Freebase\footnote{https://developers.google.com/freebase/} as knowledge base is that most of the Wikipedia page titles (or a sub-string of the title), as well as table cell entities, can be mapped to unique machine identifier (mid) in Freebase.%\looseness=-1

\subsection{Data Filtering}
\label{subsection:datacollection}

\begin{figure}[!htb]
\includegraphics[width=\linewidth]{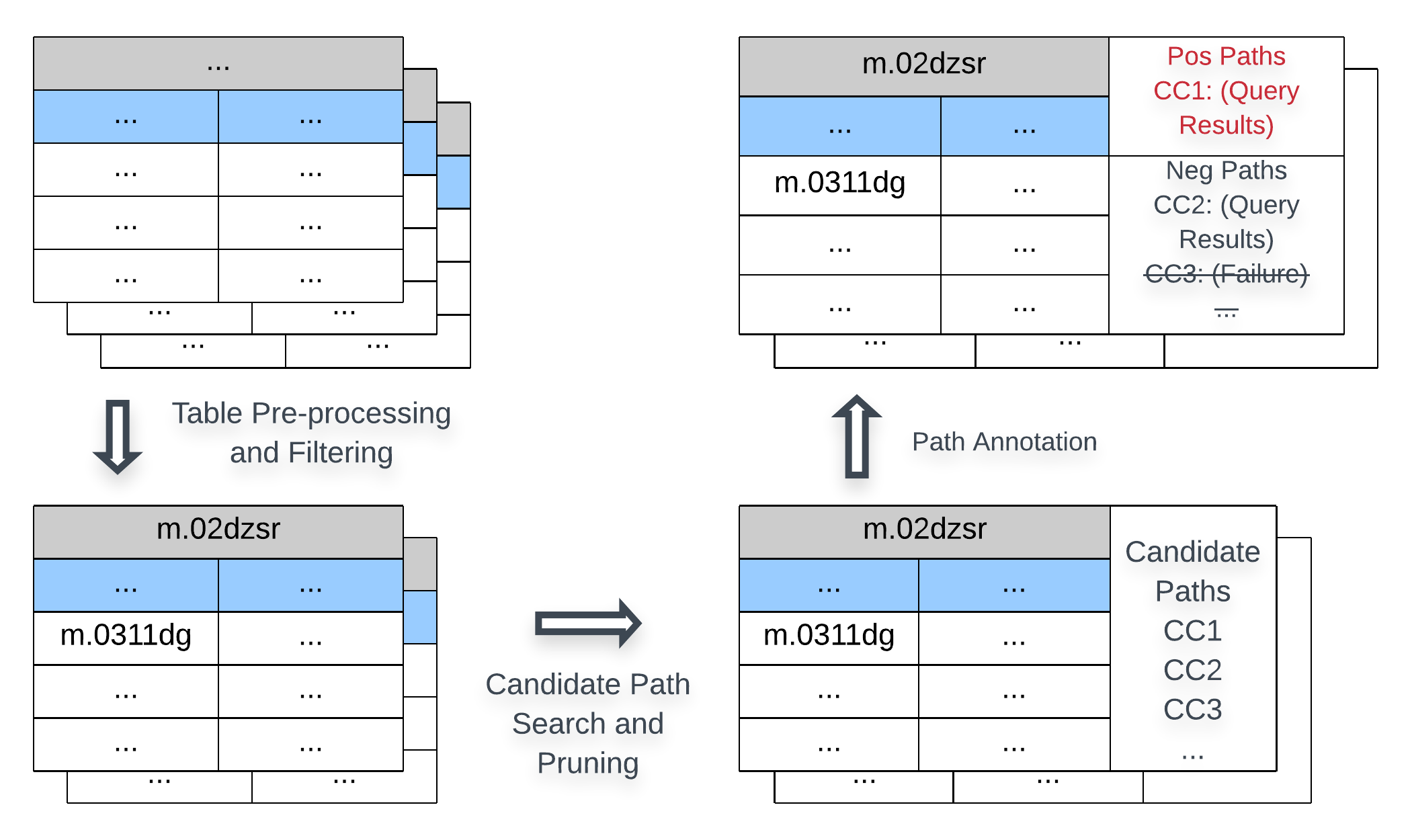}
\vspace{-20pt}
\caption{Data collection \& annotation steps.}
\vspace{-10pt}
\label{fig:datacollsteps}
\end{figure}

Our data construction procedure is summarized in Figure~\ref{fig:datacollsteps}.
The key steps of data collection are: 1) pre-processing and filtering an initial subset of tables from the 1.65M tables of WikiTable; 2) building a candidate path set for each table in the selected subset using path search and pruning; 3) path annotation and final table selection. \looseness=-1

\subsubsection{Table Pre-processing and Filtering}
\label{subsubsection:datacleaning}

We process the tables as follows: For each cell, we find hyperlinks in it and link them to Freebase based on the mapping between Wikipedia URL and Freebase entity mid. %\textcolor{blue}{([BB] XD: wasn't this already done as part of the original data ?)}\xd{This is not in the original WikiTable Data, this is done by Yao in clean\_data.json, I think this is part of our processing}
We use the first hyperlink appearing in the cell and remove the entire row if the cell cannot be linked. 
The linked entity tuples act as the desired output Result Rows (RR). To simplify the problem, we only use 2-column tables, with the leftmost column as core column and one extra column, thereby significantly reducing the table set. We adopt simple rules for core column detection~\cite{venetis2011recovering, zhang2017effective, wang2012understanding}: (i) the core column is the leftmost column; (ii) all linked entities in the core column are unique.
We then normalize Column Names by lemmatizing the strings into nouns.
We remove tables with no core column, empty column names or if the table has less than 3 linked \textbf{$RR$}.
The last rule of atleast 3 KB-linked $RR$ helps us in determining positive paths in the later stage of path search and annotation.\looseness=-1

The next step is to extract the \textbf{Subject Entity (SE)} from the Page Title string for the selected set of tables.
We do not use string based mapping of page title to entities, because some page title does not have direct mapping as the entity is included inside the title string (e.g., List of \textsf{Entity X}).
We use Google Knowledge Graph Search API\footnote{https://developers.google.com/knowledge-graph/} to get SE by querying with page title string and extracting the mid from the \textit{top-1 retrieved result}.
If the API returns an entity name for the top-1 \textit{mid}, we directly use it as EN, otherwise, we retrieve the name from Freebase using the entity \textit{mid}.
Tables with no matching entity linking for title or with empty EN for the linked SE \textit{mid} are discarded.
The result of entity linking is then evaluated by string matching between the selected (top-1) entity name and Page Title string using a simple heuristic: we check whether the extracted entity name is in the Page Title string or vice-versa, and remove a table if it does not pass this check.
The concatenation of Page Title with Table Caption, after removing EN, constitutes the \textbf{Query Intent String (QIS)} for each table.\looseness=-1

We gather the \textbf{Subject Entity Types (SET)} of the extracted \textbf{SE} mid, by concatenating the least frequent Freebase type with the fine-grained type system built by \cite{ling2012fine}, when a mapping of Freebase type to fine types exist.\footnote{We prune too generic types starting with prefix: [base, common, type]}
This can be thought of as an additional normalization of Freebase types (e.g. tennis\_tournament, formula\_1\_grand\_prix, and olympic\_games are all mapped to sports\_event) and helps capture more general type information of the subject entity, along with the more specific sub-type information from Freebase. We perform basic string pre-processing on text fields to replace number and empty string with two special tokens, \textit{numtkn} and \textit{emptstr}, respectively.\looseness=-1

\subsubsection{Candidate Path Search and Pruning}
\label{subsubsection:pathfinding}

For each table with SE gathered above, our next task is to create a set of candidate $[P_1 - P_2]$ chains, such that when converted to SPARQL queries, these chains can retrieve a subset (or all) of the entity tuples (rows) of the original table.
The \textit{Max\_Recall} of each $[P_1 - P_2]$ is essentially the proportion of entity tuples (rows) of that table retrieved by it.
For each table with a SE, for each input row $(ER_1, ER_2)$, where $ER_1$ is the column 1 entity, and $ER_2$ is the column 2 entity of the row $ER$, we search for candidate $P_1$ paths between $(SE, ER_1)$ and candidate $P_2$ paths between $(ER_1, ER_2)$ in Freebase, with an additional constraint that the maximum number of sub-paths (hops) for any candidate path in $CC_1$ and $CC_2$ is atmost 3.\looseness=-1

%\ttfamily,
\begin{lstlisting}[basicstyle=\small,language=SPARQL]
prefix : <http://rdf.basekb.com/ns/>    
SELECT * WHERE
{{?s0 ?p0 :e_0. 
FILTER (isIRI(?s0) && 
?p0 != rdf:type && STRSTARTS(STR(?s0), STR(:))).}
UNION
{:e_0 ?p1 ?s1. 
FILTER (isIRI(?s1) && 
?p1 != rdf:type && STRSTARTS(STR(?s1), STR(:))).}}
\end{lstlisting}

To simplify our path search, we first retrieve from Freebase the 2-hop neighborhood i.e., relationship and neighboring entities, of each of the unique $SE$, $ER_1$ and $ER_2$ entities present in our collected tables.
We execute a simple SPARQL query (shown next) on each such unique entity $e\_0$, and then on its collected neighbors, with the additional constraint that we do not expand nodes with more than 500 neighbors.\looseness=-1

The query considers both incoming and outgoing relationship edges for each entity $e\_0$ in Freebase through the paths $p0$ and $p1$ respectively.
We store the neighbors as $(e\_0,$ \textasciicircum $p0, s0)$ and $(e\_0, p1, s1)$, where \textasciicircum \;is used to represent the  inverse of a relationship during SPARQL query execution, such that the relationship edges now implicitly encode the direction information as part of the modified label.
This allows us to construct undirected subgraphs between any two entity pairs locally using the above neighborhood information that we use to find simple paths between entity pairs.\looseness=-1

For each table with subject entity SE, for each of its row $(ER_1, ER_2)$, we create a query-able set of entity pairs using $<$SE, $ER_1$ $>$ and $<$$ER_1, ER_2$$>$ and collect these pairs in a global set $G$ of entity pairs across all tables.
For each pair $(e_i, e_j)$ in $G$, we use the neighborhood information of the entities, gathered in the prior step, to construct an undirected graph, on which we run Depth First Search (DFS) to generate candidate sequence of intermediate entities that connect $(e_i, e_j)$ in Freebase.
We set max depth for DFS, such that the maximum length of a path connecting any $(e_i, e_j)$ can be 3 i.e., total 3 edges, which implies at most 2 intermediate entities.
As we do not expand entities with more than 500 neighbors, paths that pass through such nodes will be pruned, leading to possibly longer length paths or missing paths between $(e_i, e_j)$.
Note that paths that go through hub nodes are often noisy, retrieving a very large candidate tuple set and hence ideal candidates for pruning~\cite{Gu2019}.
%DFS is a better choice over BFS due to very high degree of certain \textit{popular nodes} (eg: USA), which can cause abruptly high memory usage during local graph construction and search step.
It is possible that some entities are connected by a simple path of length $> 3$, which we do not explore due to the path length limit.
Also, some nodes in Freebase can be totally disconnected i.e, no adjacent neighbors (eg: m.012m5r2l [Stranger in the House]), and hence we cannot find any paths for such entities.\looseness=-1

Our actual candidate set of paths for $(e_i, e_j)$ is a set of all possible path combinations using all the unique interconnecting relationships between the intermediate entities (obtained by DFS above) connecting with $e_i$ and $e_j$.
This makes our candidate path set between any two entity pairs very skewed in size, depending on the set of unique neighboring relationship edges.
To reduce the set, we prune all paths that start with prefixes that are too generic/common\footnote{freebase, common.topic.notable, common.topic.image,  common.topic.webpage, type.content,type.object, dataworld.gardening\_hint}, as they are more likely to be noisy paths and semantically less related to the actual relationship between entities.
At the end of this step, for each table we have $M$ candidate $P_1$ paths obtained by path search between all $<$$SE, ER_1^i$$>$ and $N$ candidate $P_2$ paths obtained by path search between all $<$$ER_1^i, ER_2^i$$>$ using KB, where i = 1 to $|RR|$.\looseness=-1

\subsubsection{Path Annotation for Final Table Set Selection}
\label{subsubsection:pathcleaning}

We remove all tables that have either $M = 0$ or $N = 0$, and then for each table, sort the $P_1$ and $P_2$ chains in descending order of their \textit{proxy recall} i.e., number of ground truth entities that they can retrieve.
We perform a $[M * N]$ join of the candidate paths of the table while checking whether both the $P_1$ and $P_2$ chains retrieve common $C1$ item(s), to generate the final set of candidate $[P_1 - P_2]$ chains for the table.
As we consider paths with length up to 3 for $P_1$ and $P_2$ individually, the maximum total $[P_1 - P_2]$ chain length can be 6.
The $[M * N]$ join of the candidate paths can yield a very large candidate chain set for some tables.
The execution time of actual SPARQL queries, synthesized from such chains, can significantly vary due to the above reason.
Additionally, some noisy and long chains in the candidate set can cause an explosion of retrieved result rows, thereby significantly degrading the overall precision.\looseness=-1

To ensure that each table is answerable through KB within a stipulated time and produce results with certain lower bound on Recall and Precision, we synthesize SPARQL queries (shown in Section~\ref{subsection:qts}) from each candidate $[P_1 - P_2]$ chain, and execute them against Freebase, with max query time set to 120 secs and max result rows set to 10,000.
These parameter configuration makes our framework practical i.e., all our candidate template queries finish within finite time (upper bounded to 120s) and indirectly ensures that the precision is not poor (lower bounded to  0.0002) for any query.\looseness=-1
%$2/10,000$

For each chain that finishes execution without violating the Freebase constraints, we compare the set of retrieved tuples against the ground truth rows of the Table to obtain \textit{Table\_Recall}, \textit{Table\_Precision} and \textit{Table\_F1}.
%For brevity, we define \textit{Table\_Recall} here.
These $[P_1 - P_2]$ chains are sorted in descending order of \textit{Table\_Recall}, \textit{Total Path Length}, and \textit{Table\_F1}.
%\begin{equation}
%    Table\_Recall = \frac{\text{Total Correct %Rows the chain can retrieve}}{\text{Total Ground %Truth Rows in the table}} \label{eq:recall@t}
%\end{equation}
All chains with the highest \textit{Table\_Recall}, shortest total length, and high \textit{Table\_F1} are annotated as positive chains, while the remaining chains are all annotated as negative chains.
Chains for which the SPARQL queries did not finish are removed directly from the candidate chain set, as these queries violate our framework's practicality assumptions.
Additionally, all chains that cannot retrieve more than 1 ground truth rows are also removed.
These steps cause the removal of additional tables, whose candidate chain set becomes empty.\looseness=-1

%Rows = 3.0, 6.0, 11.0,  19.70, 23.0, 1062
%Pos = 1.0 & 1.0 & 1.0 & 1.26 & 1.0 & 90.0
%Neg = 0.0 & 4.0 & 12.0 & 57.53 & 39.0 & 5163

\subsection{Data Characteristics}

After all the data pre-processing steps, we are left with total \textit{4,013} tables that satisfy our defined properties.
The \textit{(min, 25-ile, 50-ile, mean, 75-ile, max)}\footnote{where \textit{x-ile} is \textit{x} percentile of the list of values} for the number of rows per table is (3.0, 6.0, 11.0,  19.70, 23.0, 1062), with about 31.85\% tables having more rows than the mean.
The \textit{median number of positive paths} per table is 1.0 while the \textit{mean} is 1.26, with maximum being 90.
The \textit{(min, 25-ile, 50-ile, mean, 75-ile, max)} for the number of negative paths per table is (0.0, 4.0, 12.0, 57.53, 39.0, 5163).
In total, there are 1749 unique positive paths, 122607 unique negative paths, of which 1042 paths are both positive and negative across all tables in our entire data set.
The skew in the number of negative paths as well as the overlap of labels for several positive paths that make our task for the QTS module particularly challenging.
We observe that $\approx$ 48.37\% of total tables have positive chain Recall $>=$ Mean Recall (0.5419), while about 9.82\% of total tables have positive chain Recall = 1.0, which means that the positive chains extracted using our strategy are meaningful and can be used for this table completion task.
Our data covers a wide range of topics represented by the diverse set of fine-grained subject entity types (SET) of many different entities (including different types of sports events, artist, etc.). Analyzing various data statistics shows that the collected dataset is both diverse and challenging. Refer to Appendix~\ref{subsection:appendixondata} for more details on the various characteristics of our dataset.\looseness=-1

\subsection{Data Partitioning}
\label{subsection:datapartitioning}

We get 4013 tables in total after data cleaning, which we divide into 80\%/10\%/10\% random split to create the training, validation, and testing sets respectively with 3209/402/402 tables.
The (mean, max) of total rows per table in Validation and Testing data are (20.83, 460) and (19.64, 342) respectively, thereby making these tables challenging to fill.
The (mean, max) total paths (positive and negative combined) per table in Validation and Testing data are (68.03, 2511) and (59.53, 2748) respectively, which makes the evaluation of chain selection performance of QTS realistic.

For training data, we pair each positive path with at least $(k-1)$ negative paths (with $k$ = 10). In case there are less than $(k-1)$ negative paths for a table, we randomly sample paths that appear as negative paths in our data set, and add them as negative paths for the table till it has at least $(k-1)$ negative paths.
Note that for tables in validation and testing set, we do not add any extra negative paths other than what it originally contains, even if the table contains less than $(k - 1)$ negative paths or zero negative paths.
Due to this reason, 24 tables (5.97\%) in validation and 23 tables (5.72\%) in testing set have only positive paths, but no negative paths at all.
We collect the set of unique words from tables in training and validation set (after removing stop words) to create Table information vocabulary (TB\_Vocab) and KB information vocabulary (KB\_Vocab).
TB\_Vocab is constructed using all words in QIS and CN fields, while KB\_Vocab is constructed using all words in SET and CC fields, filtering out words that appear only once. We introduce a special Out-of-Vocabulary (OOV) token to replace all words that are not retained in the corresponding vocabulary.

\section{Experimental Setup}
\label{section:experimentalsetup}

We implemented all components of our framework in Python. We use the same Training/Validation/Testing split across all experiments. The BOW based models and the DNN model for QTS have been implemented using scikit-learn~\cite{scikitlearn} and Tensorflow~\cite{tensorflow2015whitepaper} respectively. We set up Freebase server with Virtuoso having 2 * E5-2620 v4 with 256GB RAM, ServerThreads = 30, MaxQueryMem = 30G, MaxStaticCursorRows = 10000, ResultSetMaxRows = 10000, MaxQueryCostEstimationTime = 300 and MaxQueryExecutionTime = 120. Experiments are run on 96 GB memory machine with 20 cores in a node hosted at Ohio Supercomputer Center~\cite{Pitzer2018}.\looseness=-1

%All experiments, including model training and end-to-end (E2E) scenarios, are run using 40 cores and 192 GB memory in a node hosted at Ohio Supercomputer Center~\cite{Pitzer2018}.

\subsection{Model Training}
\label{subsection:modeltraining}

\subsubsection{Query Template Selector}
\label{subsubsection:qtslearning}

The learning sub-component of our Query Template Selector module is a matching framework, which scores each candidate chain given the query intent for each table, and then this score is used to select the most matching chain.
We have described a DNN model (in Section~\ref{subsection:qts}) as one instance of such a learning sub-component.
In order to compare the performance of DNN model for chain selection, we construct several alternative models (both non-learning and learning based) viz: Random, JacSim, KNN, LR, RF and use them as a scoring strategy for chain selection.\looseness=-1\vspace{-0.2cm}
\begin{itemize}
    \item \textbf{Random:} We concatenate and shuffle the positive and negative chains of each table and then randomly sample a chain from this set.
    \item \textbf{Unsupervised non-learning based scoring strategy:} We use the set of $<$QD, SET, CN$>$ tokens and the set of tokens from a candidate chain to compute the set overlap based Jaccard similarity (\textbf{JacSim}) that is used as a matching score to rank candidate chains.
    \item \textbf{Supervised scoring strategy:} We use Training data as input to the K-Nearest Neighbor (\textbf{KNN}), Logistic Regression (\textbf{LR}) and Random Forest~\cite{scikitlearn} (\textbf{RF}) to train a binary classifier, and then use the trained model to generate matching scores.
\end{itemize}

\textbf{\noindent Input:} For KNN, LR and RF, we use $<$QD, SET, CN$>$, as well as positive and negative candidate chains ($CC$) as inputs to the models. We use Table\_Vocab to fit 3 CountVectorizer~\cite{scikitlearn}, one each for QD, CN1 and CN2, while we use KB\_Vocab to fit 2 CountVectorizer~\cite{scikitlearn}, one each for SET and CC.
For each table, we concatenate all these 5 vectors, the variable one being the vector for unique chain from $CC$.
Depending on whether the positive or negative chain is picked from $CC$, we label the concatenated vector as positive (+1) or negative (-1) respectively.
%\st{the \textit{predict\_proba()}
\begin{comment}
\begin{table}[!htb]
\small
\centering
\begin{tabular}{|p{1cm}|p{6cm}|}
\hline
Model & Parameter Grid\\
\hline
KNN & n\_neighbors : [5, 10, 25, 50, \textbf{100}, 250, 500], weights : [\textbf{`distance'}] \\
\hline
LR &  C: np.power(10.0, np.arange(-5, 5)) [\textbf{0.001}], \newline
        solver: [\textbf{sag}, saga], \newline      max\_iter: [1000, \textbf{2500}, 5000, 10000] \\
\hline
RF &   n\_estimators : [50, 100, \textbf{250}, 500], \newline
max\_depth : [8, 16, \textbf{32}, 64, 128, 256], \newline
criterion : [\textbf{entropy}], \newline
class\_weight:[balanced,\textbf{balanced\_subsample}]\\
\hline
\end{tabular}
\caption{Grid search parameters for each model. The best parameters obtained are highlighted.}
\vspace{-20pt}
\label{tab:gridsearchparams}
\end{table}
\end{comment}

\textbf{\noindent Training:} The first step is to perform Grid Search for optimal parameters for KNN, LR and RF models using scikit-learn's GridSearchCV module~\cite{scikitlearn}. 
For this, we concatenate Training and Validation data and run 3 fold cross-validation for each of the models to find optimal parameter settings using \textit{roc\_auc} for scoring.
Once the best parameter configurations are found for each model, we initialize respective models using the corresponding best parameters and then train each model from scratch using only the Training data.
%from the corresponding parameter grid (Table~\ref{tab:gridsearchparams}) 

\textbf{\noindent Evaluation:} For each table in the Validation/Testing data, we use all available chains as candidates and create the input data array $X$ by concatenating the query intent with each of the candidate chains to generate input data array (as discussed in Input part above) using the trained Vectorizers.
We use the trained model (fitted with best parameters as described above) to retrieve the \textit{positive class score} (score for label +1) returned by the above methods for each row in the input data.
These scores are used to sort the candidate chains in descending order to pick the top-1 chain as the predicted chain, and compared with the ground truth positive chain of the corresponding table to compute the Accuracy@1. (results in Section~\ref{subsection:queryselectorerror})\looseness=-1

%\st{then invoke the \textit{predict\_proba($X$)}~\cite{scikitlearn} on} \fromH{because "\textit{predict\_proba($X$)" is not informative}} 
%[BB] I had cited scikit learn 

\textbf{\noindent DNN Model:} 
We initialize the individual components of our DNN model (proposed in Section~\ref{subsection:qts}) as: a) Intent String Encoder (ISE):~\footnote{settings = output encoded vector size - input token embedding size - filter1 size,filter2 size,filter3 size} $settings = 100-100-1,2,3$ with \textit{maximum token length = 100}; b) Header String Encoder (HSE): $settings = 25-100-1,2,3$ with \textit{maximum token length = 10}; c) Entity Type Encoder (ETE): $settings = 100-100-1,3,5$ with \textit{maximum token length = 100}; d) Path Chain Encoder (PCE): $settings = 250-100-1,3,5$ with \textit{maximum token length = 200}.
We train our model on the training data using \textit{$9$ negative chains per table} in a mini-batch.
We use Adam Optimizer with \textit{learning rate = 0.00001}, \textit{$\lambda$ = 0.000005}, \textit{m = 0.25} and train for \textit{2000 epochs} with \textit{mini batch size = 250}.
All token embeddings are learned from scratch and are used during Validation/Testing along with the weights of the trained model.\looseness=-1\footnote{The vocabulary for ETE and PCE are the same, and we initialize and learn common embedding matrices for these encoders.
We initialize and learn separate word embeddings for ISE and HSE, although they have the same vocabulary.}

\subsubsection{Candidate Tuple Ranker}
\label{subsubsection:ctrlearning}
We train the ranker using LambdaMART algorithm \cite{burges2010ranknet} with the implementation of XGBoost \cite{chen2016xgboost}. To generate training data, for each table we use the first positive chain and a row associated with that chain as an example row.
We run grid search using the concatenated training and validation data, and then refit the model with the best-chosen parameters using only the training data.\looseness=-1

\subsection{KBTableFill using Trained Modules}
\label{subsection:e2escenario}

\begin{algorithm}[!htb]
\caption{End-To-End Scenario}
\label{alg:e2escenario}  
\begin{algorithmic}[1]          
\REQUIRE Pre-trained QTS model; Pre-trained CTR model; 
\REQUIRE Pre-processed Tables (Val/Test) set $T$
%\STATE min\_Recall = 0.0, av\_Recall = 0.0, max\_Recall = 0.0
%\STATE min\_Recall@R = 0.0, av\_Recall@R = 0.0, max\_Recall@R = 0.0
\FOR {Table $t \in T$}
  \STATE Retrieve pre-processed $<$QIS, CN, SE, SET, CC$>$
  \STATE Retrieve $RR$ as list of ground truth tuples of $t$
  \STATE Compute number of GT rows of $t$: $R = |RR|$
  \FOR {$ER$ $\in$ $RR$} 
      \STATE Remove $ER$ from $RR$ to build $ERR$
      \STATE Use $<$SE, ER$>$ to filter $CC_{ER}$  from $CC$
      \STATE Send $<$QIS, CN, SET, $CC_{ER}$ $>$ to pre-trained QTS to compute matching score, rank the chains in descending order of score and pick top-1 chain $BP$\looseness=-1
      \STATE Use $SE$ and $BP$ to synthesize SPARQL query $SQ$
      \STATE Execute $SQ$ to retrieve candidate tuples ($CT$)
      \STATE Use $CT$ and $ERR$ to compute Tuple\_Recall
      \STATE Get $CT$ features for $<$QIS,SE,SET,CN,BP,ER$>$
      \STATE Use features to rank $CT$ using pre-trained CTR
      \STATE Compute NDCG using ranked $CT$
  \ENDFOR
  \ENDFOR
\RETURN
\end{algorithmic}
\end{algorithm}
%\vspace{-0.25cm}

In the end-to-end (E2E) scenario, the user input tabular query, after requisite pre-processing by the Query Pre-processor (QP) component, is fed into the Query Template Selector (QTS) component and the resulting table is produced as output of the Candidate Tuple Ranker (CTR) component. For experiment purpose, we assume that the pre-processing of the input data has been completed offline, which allows us to directly construct the pre-processed output of tabular query, which is supposed to be generated by QP.
We individually train the QTS module and CTR module first (as described in previous sections), and then use the trained modules for E2E.\looseness=-1

For each table $t$ in the Validation/Testing set, containing $R$ rows (where $R = |RR|$), we simulate our framework by selecting each result row ($RR$) as proxy for user example row ($ER$) and run it through the end-to-end flow, as described in Algorithm~\ref{alg:e2escenario}. For each table $t$, given SET and current input row $(ER_1, ER_2)$, we use the set of candidate chains of that table $CC_{ER}$ present in the dataset, with the additional constraint that each such chain has to connect the current tuple $(ER_1, ER_2)$ in KB via a valid path.
Thus $CC_{ER}$ is generally a subset of the overall candidate chains $CC$ of the whole table.
%The QTS module assigns a matching score to each chain based on the query context using the specific model used.The candidate chains are sorted in descending order of computed scores, and the top-1 chain ($BP$) is picked as the prediction by this module.We use $SE$ and $BP$ to synthesize the SPARQL query, execute it and retrieve a set of candidate result tuples (CT), which are used in conjunction with $ER$ and $BP$ to perform feature driven ranking of the candidate tuples using the trained CTR module.
It is possible that for some $ER$, the $CC_{ER}$ set is empty i.e., no valid path is found between $ER_1$ and $ER_2$ that would satisfy all our constraints during the data gathering phase.
Simulations on such candidates are skipped.
For each table, Expected Result Rows ($ERR$) is constructed in every iteration, by removing the selected $ER$ from the ground truth result rows, which makes the cardinality of $ERR$ as $(R-1)$. 
\looseness=-1
%\vspace{-0.2cm}
%We compute Tuple\_Recall using expected result rows ($ERR$) as ground truth rows, and not the whole $RR$, which are used to measure the qualitative performance of our framework.
%This makes Tuple\_Recall a bit different from Table\_Recall, which is computed on $RR$ instead of $ERR$.

\subsection{Evaluation Strategy}
\label{subsection:evaluation}

%\begin{equation}
%    Tuple\_Recall = \frac{\text{Total Correct %Result Rows (removing ER)}}{\text{Total Rows in %Ground Truth  (removing ER)}} %\label{eq:recall@er}
%\end{equation}

We use Tuple\_Recall as the primary measure of performance of \tabfill for the E2E scenario.
Tuple\_Recall computes how many rows of the original table (except the ER) can be retrieved by the chain obtained using current ER.
%Since in E2E simulation, we provide an Example Row (ER), we explicitly remove it from the set of tuples retrieved by any chain in $CC_{ER}$, as all chains in $CC_{ER}$ will always retrieve ER, and hence the results will be biased otherwise.
%On the contrary, Table\_Recall of any chain gives us an estimate of how many rows of the table can be retrieved by it.
For CTR, we use NDCG with a binary relevance score, where a correct tuple has a score of 1 while an incorrect tuple has 0, as well as Precision@1 of the ranked result rows, after removing ER.\looseness=-1

\subsection{Core Column Entity Retrieval Baseline}
\label{subsection:corecolbaseline}

One of the tasks in~\cite{ZhangSmartTable} is to retrieve candidate entities for the core column (similar to C1 column in our setting), given a partially completed table as input.
Hence, the technique proposed by~\cite{ZhangSmartTable} is used as baseline to evaluate the performance of KB meta-path based entity retrieval strategy. The authors use both Wiki Tables and Knowledge Base as information source. To use Wiki Tables, they first retrieve top-k most similar tables based on table caption and seed entities, all the entities in these tables are then taken as candidates. To use Knowledge Base, they simply select top-k entity types based on seed entities and use all entities having these types as candidates. Following~\cite{ZhangSmartTable}, we set k as 256 for tables and 4096 for types.
To make the model applicable to our settings, the following changes have been made:
1) The table corpus released by~\cite{ZhangSmartTable} is used, but all DBpedia entities are mapped to Freebase entities and all tables that appear in our validation/testing set are removed, 2) we use Freebase RDF types instead of DBpedia types. \looseness=-1
%More details of the baseline can be found in Section 4.1 of ~\cite{ZhangSmartTable}.

%to the publicly available code\footnote{https://github.com/iai-group/sigir2017-table}
\section{Experimental Results}
\label{section:experiments}

\begin{table*}[!htb]
\centering
\large
\resizebox{\textwidth}{!}{\begin{tabular}{|p{2cm}|c|c|c|c|c|c|}
\cline{1-7}
Scenario \newline [QTS, CTR] & \multicolumn{3}{|c|}{\bf{Validation Set (Tabular Queries = 4753)}} & \multicolumn{3}{|c|}{\bf{Testing Set (Tabular Queries = 4191)}}\\
\cline{2-7}
& \multicolumn{1}{|c|}{\bf{Tuple\_Recall}} & \multicolumn{1}{|c|}{\bf{NDCG@All}}
& \multicolumn{1}{|c|}{\bf{P@1}} & \multicolumn{1}{|c|}{\bf{Tuple\_Recall}}
& \multicolumn{1}{|c|}{\bf{NDCG@All}} & \multicolumn{1}{|c|}{\bf{P@1}}\\
\cline{2-7}
& [25-ile,50-ile, \textbf{Mean},75-ile] & [25-ile, \textbf{Mean},75-ile] & [\textbf{Mean}] & [25-ile,50-ile, \textbf{Mean},75-ile] & [25-ile,\textbf{Mean},75-ile] & [\textbf{Mean}]\\
\cline{1-7}
Rand, Rand & 0.1111, 0.3333, \textbf{0.3873}, 0.610 & 0.1781, \textbf{0.3444}, 0.4336 & \textbf{0.0970} & 0.1024, 0.3103, \textbf{0.3628}, 0.5714 & 0.1615, \textbf{0.2925}, 0.4136 & \textbf{0.0465}\\
\hline 
\rowcolor{Gray}
Rand, FR & 0.1111, 0.3333, \textbf{0.3853}, 0.625 & 0.2218, \textbf{0.4310}, 0.6048 & \textbf{0.2125} &  0.1048, 0.3125, \textbf{0.3638}, 0.5698 & 0.1982, \textbf{0.3794}, 0.5389 & \textbf{0.1510}\\
\hline 
%LR, Rand & 0.2143, 0.4667, \textbf{0.4804}, 0.8 & 0.2242, \textbf{0.3995}, 0.4784 & \textbf{0.1462} & 0.1847, 0.4286, \textbf{0.4596}, 0.7143 & 0.1955, \textbf{0.3411}, 0.5010 & \textbf{0.0656} \\
%\hline 
%\rowcolor{Gray}
%LR, FR  & , , \textbf{}, & , \textbf{}, &  \textbf{} & , , \textbf{}, & , \textbf{}, & \textbf{} \\
%\hline 
%KNN, Rand & 0.2308, 0.5, \textbf{0.490956}, 0.8 & 0.2201, \textbf{0.3923}, 0.4704 & \textbf{0.1368} & 0.1994, 0.46, \textbf{0.4753}, 0.75 & 0.1981, \textbf{0.3318}, 0.4867 & \textbf{0.0570} \\
%\hline 
%\rowcolor{Gray}
%KNN, FR  & 0.2308, 0.5, \textbf{0.490956}, 0.8 & 0.2802, \textbf{0.4811}, 0.6735 &  \textbf{0.2527} & 0.1994, 0.46, \textbf{0.4753}, 0.75 & 0.2440, \textbf{0.4262}, 0.5688 & \textbf{0.1854} \\
%\hline 
RF, Rand & 0.2623, 0.5, \textbf{0.5163}, 0.8126 & 0.2245, \textbf{0.3880}, 0.4612 & \textbf{0.1288} & 0.2128, 0.4590, \textbf{0.4832}, 0.75 & 0.1996, \textbf{0.3339}, 0.4853 & \textbf{0.0530}\\
\hline 
\rowcolor{Gray}
RF, FR & 0.2623, 0.5, \textbf{0.5163}, 0.8126 & 0.2891, \textbf{0.4830}, 0.6576 & \textbf{0.2537} & 0.2128, 0.4590, \textbf{0.4832}, 0.75 & 0.2471, \textbf{0.4274}, 0.5693 & \textbf{0.1813}\\
\hline 
DNN, Rand & 0.25, 0.5, \textbf{0.5099}, 0.8391 & 0.2186, \textbf{0.3848}, 0.4696 & \textbf{0.1195} & 0.2047, 0.4545, \textbf{0.4754}, 0.75 & 0.1919, \textbf{0.3299}, 0.4890 & \textbf{0.0582}\\
\hline 
\rowcolor{Gray}
DNN, FR & 0.25, 0.5, \textbf{0.5099}, 0.8391 & 0.2750, \textbf{0.4736}, 0.6509 & \textbf{0.2362} & 0.2047, 0.4545, \textbf{0.4754}, 0.75 & 0.2390, \textbf{0.4169}, 0.5622 & \textbf{0.1696} \\
\hline 
\end{tabular}}
\caption{E2E scenario results for \textit{tabular queries} that successfully executed across all tables in the respective partitions. Each scenario is a unique combination of QTS and CTR sub-modules that have been stitched together for the E2E simulation using Algorithm~\ref{alg:e2escenario}. The number of successful queries is reported at the top. There is a clear performance improvement for both modules when learning based sub-components are used. [RF, FR] and [DNN, FR] are the top 2 best performing scenarios, with significant qualitative improvement of the Random baseline. \looseness=-1}
\vspace{-20pt}
\label{tab:e2escenario}
\end{table*}

%[Validation] Total Tables with E2E simulations : 402, Total Expected Tabular queries : 8374, Total Executed Tabular queries : 4753 (56.75901600191067 %)
%[Test] Total Tables with E2E simulations : 402, Total Expected Tabular queries : 7897, Total Executed Tabular queries : 4191 (53.07078637457262 %)

\subsection{End-to-End Scenario}
\label{subsection:e2escenarioresults}

\tabfill is a modular framework, where the decision making components are QTS and CTR.
Hence for the E2E evaluation, we pick specific combinations of individual modules and simulate Algorithm~\ref{alg:e2escenario}.
We pick Random chain sampling (Rand), Random Forest model (RF), which is the best performing amongst the classifier based models (details in Table~\ref{tab:queryselection}) and the DNN model (DNN), as 3 different methods for the QTS module, while using Random shuffle (Rand) and Feature driven Ranking (FR) as 2 different methods for CTR module.
The performance of our framework is evaluated using these 6 configurations for the {tabular query resolution} task on the Validation and Testing set, by running E2E for one complete simulation.
As we use all rows of each table for simulation, there are a total 8,374 and 7,897 unique tabular queries on the Validation and Testing data respectively, out of which 4753 and 4191 are actually executed.
The remaining 3621 (43.24 \%) and 3706 (46.93 \%) queries in Validation and Testing set fail, as the $CC_{ER}$ is empty (analysis in Section~\ref{subsection:dataerror}).
Tuple\_Recall measures the performance of QTS module, while NDCG@All and P@1 are used to evaluate QTS and CTR jointly.\looseness=-1

We compute the performance numbers using results for only those queries that completed successfully under different selection of [QTS, CTR], and summarize them in Table~\ref{tab:e2escenario}. 
The minor difference in Tuple\_Recall for the Random models is an artifact of randomness in chain selection.
Both RF and DNN based scenarios improve over Random QTS by at least 0.1 in the Testing data in terms of Mean Tuple\_Recall, with RF model being marginally better than DNN model.
Also, the 75-ile of Tuple\_Recall on Validation data for DNN QTS is 0.8391, compared to 0.8126 of RF and 0.610 of Best Random strategy (in terms of Mean Tuple\_Recall), which shows that our learning models are performing consistently better than random chain selection strategy (additional details in Section~\ref{subsection:queryselectorerror}).
Precision@1 is significantly better (close to 2x improvement in Validation and close to 3x improvement in Testing data) and there is a consistent gain in NDCG@All for all models using the feature based ranker, when compared to the random shuffling baseline (additional details in Section~\ref{subsection:ctrerror}).
The performance of FR depends on the candidate tuples generated by the QTS selected chain.
Thus when QTS performs better (Tuple\_Recall of RF better than DNN), the performance of FR is marginally better.
%We observe that when QTS module performs better for RF than DNN in terms of Tuple\_Recall on both Validation and Testing data, the performance of feature ranker is slightly better for RF than DNN.
The E2E simulations demonstrate the benefits of using machine learning based components in \tabfill to complete tabular queries on our diverse, challenging and realistic data.\looseness=-1

\subsection{Core Column Entity Retrieval Scenario}
\label{subsection:corecolresults}

During E2E scenario execution of \tabfill (Section~\ref{subsection:e2escenarioresults}), for each successful tabular query we store the following metadata: table id, Subject Entity, current example row from the tabular query and the predicted chain [$P_1-P_2$] by the corresponding QTS module.
Later, for each tabular query, we pick either the $C_1$ sub-path [$P_1$] or the full path [$P_1-P_2$], and synthesize the SPARQL query to retrieve candidate $C_1$ entities directly or in the later case candidate rows (<$C_1$,$C_2$> pairs) from which the unique $C_1$ entities are extracted.
In both the scenarios, $C_1$ Recall of the result is computed using the ground-truth $C_1$ entity set for that table id, after removing the $C_1$ example entity.
For the baseline, which uses the same set of tabular queries as above, we vary the information source for candidate retrieval, thereby resulting in 3 settings viz: 1) Only Wiki Tables (WT) 2) Only Knowledge Base (KB) 3) Both KB and WT.
The results are presented in Table~\ref{tab:corecolresulttable}.\looseness=-1

\begin{table}[!htb]
\small
\centering
\begin{tabular}{|p{2cm}|p{3cm}|p{2.75cm}|}
\cline{1-3}
\textbf{Method} & \textbf{Validation $C_1$ Recall (\# Queries = 4753)} & \textbf{Testing $C_1$ Recall \newline (\# Queries = 4191)}\\
\cline{2-3}
& [50-ile, \textbf{Mean},75-ile] & [50-ile, \textbf{Mean},75-ile]\\
\cline{1-3}
B [Only WT] & 0.2375,\textbf{0.4227},0.9167 & 0.3420,\textbf{0.4369},0.8864 \\
\hline
B [Only KB] & 0.0833,\textbf{0.2624},0.4213 & 0.1111,\textbf{0.2903},0.4789 \\
\hline
B [KB \& WT] & 0.4423,\textbf{0.5070},0.9600 & 0.6130,\textbf{0.5386},0.9667 \\
\hline
\hline
\rowcolor{Gray}
TF (Rand) [$P_1$] & 0.7329,\textbf{0.6212},0.9333 & 0.6471,\textbf{0.5953},0.9042\\
\hline
TF (Rand) [Full] & 0.5185,\textbf{0.5065},0.7857 & 0.5,\textbf{0.4969},0.8\\
\hline
\rowcolor{Gray}
TF (RF) [$P_1$] & 0.8636,\textbf{0.7131},0.9826 & 0.7857,\textbf{0.6879},0.9602\\
\hline
TF (RF) [Full] & 0.7640,\textbf{0.6475},0.8699 & 0.7083,\textbf{0.6353},0.9 \\
\hline
\rowcolor{Gray}
TF (DNN) [$P_1$] & 0.8462,\textbf{0.7001},0.9714 & 0.7826,\textbf{0.6753},0.9602\\
\hline
TF (DNN) [Full] & 0.75,\textbf{0.6360},0.8889 & 0.68,\textbf{0.6189},0.9\\
\hline
\end{tabular}
\caption{Core column entity retrieval performance comparison between the feature driven baseline (B)~\cite{ZhangSmartTable} and various QTS modules of \tabfill (TF), with just [$P_1$] path or Full i.e., [$P_1-P_2$] path.\looseness=-1}
\vspace{-20pt}
\label{tab:corecolresulttable}
\end{table}

When QTS module uses learning based chain selection strategy i.e., RF and DNN, \tabfill consistently outperforms the best performing baseline (which uses both KB \& WT as information source) in terms of Mean $C_1$ Recall on both Validation and Testing data.
The performance of the non-learning based Random chain selector for \tabfill is better than the baseline for [$P_1$]  setting, but poorer for the [$P_1 - P_2$] setting.
In general, for \tabfill the Mean $C_1$ Recall is higher for [$P_1$] than [$P_1 - P_2$].
Rows (i.e.,<$C_1$,$C_2$>) with correct $C_1$ entity, are not retrieved when the $P_2$ path does not connect the $C_1$ entity (retrievable using only the $P_1$ path) with any $C_2$ entity, thereby causing a dip in Recall when full [$P_1 - P_2$] path is used.
There is a wide gap in performance between the baseline~\cite{ZhangSmartTable} and \tabfill when both use \textit{only KB} as the information source. 
It is our understanding that \tabfill performs better due to the principled KB meta-path based entity retrieval strategy compared to feature driven entity retrieval technique of the baseline.\looseness=-1

%there is no valid [$P_1 - P_2$] between <$C_1$,$C_2$>, i.e.,

\subsection{Error Analysis}
\label{subsection:errorsources}
Our framework is modular, each module having its own source of errors.
The components are invoked in sequential order, meaning that error from each component propagates to follow-up modules.
The two main subtasks that impact the Recall measure, i.e., the number of ground truth tuples retrieved as part of E2E simulation, are 1) KB linking and path composition between entities in data collection and pre-processing; 2) picking the best chain from candidate KB chains in the Query Template Selector component.  At the end, we also discuss the Candidate Tuple Ranker component. %which affects the precision related measures of the top-ranked entity tuples. 
%\vspace{-0.15cm}

\subsubsection{Data Collection \& Pre-processing}
\label{subsection:dataerror}

%We use the last Freebase dump from April 2015 while the WikiTable dataset is extracted from November 2013 English Wikipedia.
The errors are: \newline
\noindent\textbf{Entity linking related:} In the entity linking step of data collection, we use Wikipedia url of cell values in WikiTable corpus to find a matching entity in Freebase.
There are some cell values that are just strings, i.e., they do not have any Wikipedia url and hence cannot be linked to Freebase mid using our strategy.
However, it is possible that the cell value is a valid entity in Freebase, and possibly our synthesized path queries retrieve it as well, but it is not counted as part of our Recall computation, as it does not have a valid url. This can make chain labels noisy in some of the tables, flipping them from potentially positive to a negative label.\looseness=-1
%Also, for the same table cell, there can be multiple possible hyperlinks, each of which can potentially link to different mid's in KB. However we always pick the first hyperlink and the corresponding mid. (eg: composite tables, like the ones on Tony Award, that merge multiple entities in a cell, or columns that are just for comments)
%Currently our $SE$ linking from Wikipedia Page Title is very conservative as we look for exact string match.
%Relaxing this check can add more tables in our data, that we can exclusively use for training our data hungry models.

\noindent \textbf{KB incompleteness related:} Since we use KB meta-paths to represent the relationship between column entities, all tables generated by our framework have cell values that are entities with valid machine id's mapped to KB.
This design choice currently limits our framework to generate tables with only KB linked column values, but it enables us to better model the latent structural relationship between table columns for the table completion task.
Intuitively, \tabfill can model and generate all possible relationships that is contained in Freebase between any pair of entities. However, in practice, Freebase like other KBs, suffers from incompleteness problems that majorly contributes to errors.\looseness=-1

%\bb{expected simulations: Val - 4753 of total 8374 sim; Test - 4191 of total 7897 sim}
%\noindent \textbf{Meta-path length constraint related.}
%[BB] This is a mix of KB incompleteness as well as path length constraint issue
Assuming full simulations across all rows per table, about 43.24\% (46.93\%) tabular queries in Validation (Testing) data never execute, as our framework fails to find paths of length $<=$ 3 between entity pairs.
One possibility is that some nodes in Freebase can be disconnected i.e, no adjacent neighbors (eg: m.012m5r2l [Stranger in the House]), and hence we cannot find any paths for such entities.
\tabfill is incapable of filling tables if such entities appear in the tabular query or are expected as rows of the output table.
%Currently we observe that only \bb{xx queries in Val and yy queries in Test} fail due to this reason, and thus it appears to be less of a problem in the current data.
Currently a minor percentage of the failures are due to this reason, and hence, less of a problem in the current data.
Note that, we already discard tables that do not satisfy our path based constraints.
However, since the table is in our data and ER from it has been picked for simulation, it means the table contains atleast one positive path connecting minimum 2 entity pairs.
Although that positive path represents the dominant relationship between entities in that table, due to KB incompleteness that path is not present between the selected ER, causing simulation failure.\looseness=-1

%\noindent \textbf{High-degree node related.}
\noindent \textbf{Meta-path Search and Pruning related:} It is possible that some entity pairs are connected by a simple path of length $<= 3$, but it passes through a high degree node (popular entity), and hence never expanded as part of our path search strategy.
While this path pruning scheme helps us collect diverse paths, it adds less frequent paths between entity pairs in our data, some of which appears only once in the entire data. This sparsity of path distribution adversely impacts the learning of the query template selector (more on this later).
Since we limit our path length between entity pairs to be atmost 3, we miss out on paths of length $>3$ that might connect them. This problem is aggravated by the above path pruning scheme, due to which several short paths that might be passing through high degree entities, are never explored, thereby amplifying the failures in E2E scenarios.\looseness=-1
\vspace{-0.20cm}

\subsubsection{Query Template Selector}
\label{subsection:queryselectorerror}

This component adopts machine learning to select one chain from a set of often many candidates, which is then executed as SPARQL query to retrieve result tuples, and has a direct impact on the overall results of our framework. 
Thus to evaluate this component, we check it's ability to pick the best possible chain \textit{per table level}, as these modules are trained using all chains per table.
We first implement a non-learning based \textit{Oracle} chain selector that always picks the first ground truth positive chain per table, and thus has 100\% $Accuracy@1$. We use this \textit{Oracle} to provide an upper bound on the best Table\_Recall that can be achieved by a perfect machine learning chain selector, and then compare the performance of our individual strategies.\looseness=-1

We train various models (as described in Section~\ref{subsubsection:qtslearning}) and then compute $Accuracy@1$ using all chains of a table as candidates, leveraging the ground truth positive/negative path annotation done as part of data collection.
For these experiments, we remove all tables that have zero negative chains from the Validation/Testing data, so that the evaluation metrics are not biased in our favour.
Thus we run experiments on effectively 378 tables in validation and 379 tables in testing set.
The average (of 10 runs) $Accuracy@1$ for each method on the corresponding data set is shown in Table~\ref{tab:queryselection}.
There is a significant skew in the number of positive and negative paths in our data (as described in Section~\ref{subsection:datapartitioning}) that causes the Random chain selector to perform so poorly. The raw text matching based JacSim scorer improves by a few percentage over Random selector, but the difference in Recall between the Oracle and these two techniques is large enough to necessitate \textit{advanced learning models} for QTS.\looseness=-1
 (details on JacSim failures in Appendix~\ref{subsection:qtsmoduledebugging})
\begin{table}[!htb]
\centering
\begin{tabular}{|l|c|c|c|c|}
\cline{1-5}
\textbf{Method} & \multicolumn{2}{|c|}{\bf{Accuracy@1}} &  \multicolumn{2}{|c|}{\bf{Mean Table\_Recall}}\\
\cline{2-5}
& Validation & Testing & Validation & Testing\\
\cline{1-5}
Oracle & \textbf{100.0} & \textbf{100.0} & \textbf{0.5599} & \textbf{0.5282}\\
\hline
Random & 13.36 & 12.85 & 0.2782 & 0.2644\\
\hline
JacSim & 22.28 & 19.50 & 0.3166 & 0.2980\\
\hline
LR & 45.77 & 41.24 & 0.3916 & 0.3607\\ %Total validation tables processed = 378, Total test tables processed = 379
\hline
KNN & 51.53 & 48.84 & 0.4343 & 0.4124\\ %Total validation tables processed = 378, Total test tables processed = 379
\hline
RF & 55.98 & 55.36 & 0.4780 & 0.4447\\ %Total validation tables processed = 378, Total test tables processed = 379
\hline
DNN & 57.67 & 55.94 & 0.4649 & 0.4372\\
\hline
\end{tabular}
\caption{Comparing various QTS models on tables with non-zero negative chains (avg. of 10 runs). \looseness=-1}
\vspace{-20pt}
\label{tab:queryselection}
\end{table}

The learning based models i.e., KNN, LR, RF, DNN, improve significantly in terms of $Accuracy@1$ over the non-learning based methods. Of those, RF and DNN have greater than 50\% $Accuracy@1$ on both Validation and Testing set, which is more than double the values obtained by the non-learning based models (i.e., Random and JacSim). The good $Accuracy@1$ numbers reflect directly on the mean Table\_Recall obtained by the respective models on the data set. For some tables, even though the KNN, RF and DNN models select a negative path chain (thereby resulting in low Accuracy@1), the overall average Table\_Recall is still quite competitive when compared to the Oracle's value, implying that our models learn to select high Recall paths, although the label of the selected path can be negative.
The difference in Mean Table\_Recall between the Oracle and the RF (DNN) model is 0.0819 (0.095) in Validation data and 0.0835 (0.091) in Testing data.
We observe that both RF and DNN models are very close to each other in terms of their qualitative performance, but hard to improve any further.\looseness=-1
%, both of which are close to the mean Table\_Recall difference in the respective data sets.
%Note that the mean Table\_Recall difference between positive chain and best negative chain (i.e., negative path with highest Recall) for the tables with non-zero negative paths, is approximately 0.1157 in Validation data and 0.0951 in Testing data respectively. The above comparison tells that the trained RF (DNN) model picks top-1 chain (some of which are negative chains), such that the difference in recall (i.e., degradation of qualitative performance) is not too poor, when compared to the mean difference between positive and the best negative chain in the respective partition.

One of the major challenges for the learning models is the problem of dual labels for chains. In our corpus, there are 1042 chains that appear as both positive and negative, and in some cases, for tables having very similar $<$QD, CN$>$. This can confuse machine learning models and results essentially from KB incompleteness: a valid chain, best representing the relation in $<$QD, CN$>$, may not exist between enough entity pairs in the table, as that chain does not connect some of the entity pairs in Freebase. Thus the Table\_Recall of that chain reduces and thereby the chain becomes a negative chain in that table. However that chain can still be a positive chain in another table, which has many (similarly related) entities unaffected by KB incompleteness.
Note that the entity linking issue, discussed previously, also contributes to the problem of dual labels for chains. (additional details in Appendix~\ref{subsection:qtsmoduledebugging}.)\looseness=-1\vspace{-0.25cm}
%To add to the challenges, there is a significant skew in the ratio of positive and negative paths (details in Section~\ref{subsection:datapartitioning}), along with the very high number of unique paths that appear only once in the overall data (e.g., 93,496 paths out of the total 123,314 (75.8\%) occur only once).

\subsubsection{Candidate Tuple Ranker}
\label{subsection:ctrerror}

One consistent trend we observe is that for most features, the score associated with the $C_2$ entities often has higher feature importance than the corresponding score for $C_1$ entity.
This is possibly because the SPARQL queries, generated from the best predicted chain, often retrieve highly relevant $C_1$ entities but noisy $C_2$ entities. Hence the ranker uses $C_2$ scores to determine the overall rank of a candidate entity pair.
%We observe that \textit{Pairwise Column Name and Type Match Score} is a very important feature for both C1 and C2, with C2 feature importance being much higher than corresponding C1 importance.
For C2, the most important features are \textit{Pairwise Entity Notable Type Match Score} and \textit{Pairwise Entity Description Match Score}.
For C1, we observe that \textit{Entity Type and Connecting Chain Type Match Score} and \textit{Pairwise Entity Description Match Score} are two very important features.\looseness=-1

The \textit{Query Intent and Entity Description Match Score} is of quite low importance, due to minimal overlap between the query intent and the text description of entities.
In future, QD can be used to extract and use common types of constraints (like Date, Location etc.)~\cite{yih2015semantic} (e.g., extracting the date ``2003" as constraint from ``2003 Open Championship") to construct more meaningful features for entity tuple ranking.
In general, all the BOW based Jaccard similarity scores appear to have higher importance than the embedding based Cosine similarity scores.
This is an important insight which can be used to remove the embedding score based features for improved running time of the CTR module in future versions.
Currently, it takes approximately 0.02 secs to construct all the features for one row (i.e., entity pair) by the CTR, given that we pre-load all relevant entity information in memory.
CTR can be further optimized by removing some low importance features (compromising on quality), while also leveraging both caching as well as thread level parallelism during feature construction, which we leave as future work.\looseness=-1

\section{Related Work}
\label{section:relatedwork}

\textbf{Searching, Mining and Generation of Tables:} Extracting and leveraging WebTables for a variety of tasks like mining, searching etc. have been an active area of research. 
One direction of work~\cite{cafarella2008webtables,cafarella2009data,nguyen2015result,venetis2011recovering,ZhangAdhocTable} focuses on matching and retrieving most relevant existing tables from various online sources, given a natural language search query. 
These works assume that all requisite tables exist in online sources like Wikipedia,  QA forums etc., and the task is to effectively match those tables with the input query using syntactic and/or semantic features. 
Yakout et al.~\cite{Yakout2012IEA} focus on filling cell values and even suggest new attributes (columns) for tables by employing a holistic table schema matching technique, but this depends on pre-existing web tables.\looseness=-1
%Huang et al.~\cite{Huang2019CFR} learns to rank order facts (columns) for entities based on query context, but depends on pre-existing information in KB, WebTables and Wikipedia infoboxes.
%However, the desired table, expressed by the specific tabular query, might not be present online.\looseness=-1

Given a natural language keyword query, Yang et al.~\cite{yang2014finding} propose efficient algorithms to extract tree patterns in KB that can be used for answering keyword queries. While their framework can generate a table as output, the schema of such a table is not user input driven.
Pimplikar and Sarawagi~\cite{Pimplikar2012ATQ} propose a novel technique to synthesize new tables with the user provided schema as well as context, by extracting and consolidating column and entity information (join like operation) from several relevant pre-existing web tables. 
While this work allows the user to control the set of output columns, the solution relies on existence of tables with relevant relationships that can be used to synthesize the desired table.
Zhang and Balog~\cite{ZhangSmartTable} propose a framework that takes the partial schema along with natural language description of the table and some example rows from the user, to fill-in additional entities in the core column~\cite{ZhangSmartTable}. 
While the input to this framework is similar to \textit{tabular query}, we focus on completing the whole table, instead of a single column.
For further automation of table generation, Zhang and Balog~\cite{Zhang2018OTG} design a feature based iterative ranking framework that takes only a natural language user query as input and automatically synthesizes the schema and content of the table, thereby giving no control over the schema to the user.\looseness=-1

Note that the above frameworks do not explicitly model the latent structural relationship between entities, which we do using Freebase relationship paths.
%Contrary to \tabfill, in \cite{ZhangSmartTable} the initial candidate entities are retrieved using type \& category match scores of the example entities with entities in KB, and using standard IR algorithm (e.g. BM25) to retrieve a set of tables containing the example entities.
Some frameworks~\cite{ZhangSmartTable,Zhang2018OTG} use WikiTable corpus and a KB as information sources to generate candidate result rows.
In contrast, we use WikiTable corpus to construct a novel data set that is used to train and evaluate \tabfill, while KB is used as the only information source to fill up cell values in table.
Sun et al.~\cite{Sun2016TCS} define the semantic relationship between entities in Web Tables by connecting them through relational chains using the column names as labels. % and then leverage such relationships between cell values of existing online tables for the question answering (QA) task.
In contrast, we use \textit{Freebase meta-paths} to define relational chains between KB linked entities, which is consistent and informative than labels used by~\cite{Sun2016TCS}.\looseness=-1
While the KB-entity linking based design currently prevents us from completing non-entity based cell values, it allows for a systematic modeling of latent structural relationship between entities.\looseness=-1

%Given an input natural language question, a deep neural model is trained to select the best matching relational chain from a set of candidate chains for the input question, which is then effectively utilized to extract the answer cells for the QA task. We adopt the idea of generating and ranking candidate chains connecting entities in WebTables from \cite{Sun2016TCS}.

\textbf{\noindent Question Answering using Knowledge Base:} Intuitively, QA can be considered a special case of Table completion, where the output is a list with single column, and (often) many rows. Much of QA research ~\cite{berant2013semantic,berant2014semantic,fader2014open,fader2013paraphrase,unger2012template,zou2014natural,yao2014information,yahya2012natural} has focused on using rich and diverse information present in knowledge bases for question answering task. Some works \cite{unger2012template,yahya2012natural} seek to translate user-provided natural language query into a more structured SPARQL query, by parsing the input question to generate template query patterns. In some cases~\cite{zou2014natural}, the question answering task is reduced to a sub-graph matching problem in the knowledge graph, while another approach~\cite{yao2014information} is to build an association between questions and answer motifs of Freebase leveraging the external guidance of web data. Hybrid approach~\cite{sun2015open} augments the vast information from web resources with the additional structural and semantic information present in Freebase for the question answering task. Recent approaches propose Deep Learning models to learn the mapping between natural language questions to corresponding answer entities~\cite{dong2015question} or knowledge base predicate sequences~\cite{yih2015semantic}.\looseness=-1 %However, none of these existing approaches can directly address tabular queries, which contain both natural language context as well as structural information.

\textbf{\noindent Path Construction on Knowledge Graphs:} Relevant meta-path search in knowledge base can be done in a variety of ways including supervised random walk based method~\cite{backstrom2011supervised}, using pre-defined meta-paths for guidance~\cite{sun2011pathsim,Shi2017PPR} or by enumerating all possible length bounded paths, often with additional ranking and pruning strategies~\cite{lao2010relational,wang2016relsim}. Given a set of example entities, Gu et al.~\cite{Gu2019} propose novel algorithms to generate diverse candidate paths, which best represent the latent relationship between entities. %We construct paths connecting entity pairs in Freebase as part of data collection step, but our problem setting is different. 
Given a \textit{tabular query} as input, we also generate a set of candidate paths using the example row, and then lever the contextual information of \textit{tabular query} to pick the best matching path from the set of candidates that is used to construct SPARQL query and retrieve additional entity tuples to fill rows of the table.\looseness=-1
%In future we seek to leverage some of the above path construction strategies to build a clean yet diverse set of path templates as part of our data collection step. 

\textbf{\noindent Entity Set Expansion:} Leveraging user provided examples to expand entity set have been extensively used as part of Query by Example~\cite{Zloof1975} or Query by Output~\cite{Tran2009} techniques for many different tasks. Broadly speaking, the querying task can be thought of as seeded or example driven querying, where the task is to understand the user's query intent based on the example tuples provided and broaden the result set by retrieving relevant tuples.
A vast body of work~\cite{zheng2017entity,Metzger2017,Zhang2017ESE,chen201833} levers knowledge graphs for this purpose.
Lim et al.~\cite{lim2013semantic} construct shortest-path distance in ontology graph based features for seed entities to train a classifier, that can retrieve additional entities.
Jayaram et al.~\cite{jayaram2015querying} propose a novel query by example framework called \textit{GQBE}, which extracts the hidden structural relationship between the user provided example tuples to build a Maximal Query Graph, and uses this graph to generate different possible motifs in the query lattice (answer space modelling), followed by tuple retrieval and ranking.
Wang et al.~\cite{wang2015concept} solve the concept expansion problem, where given the text description of a concept and some seed entities belonging to it, the framework outputs a ranked list of entities for that concept using pre-existing web tables.
Zhang et al.~\cite{Zhang2017ESE} lever common semantic features of seed entities to design a ranking framework to retrieve and rank relevant answer entities.
\tabfill develops over the entity set expansion paradigm, by additionally incorporating contextual and structural information (captured using KB metapath) present in the input \textit{tabular query} to retrieve and rank \textit{list of entity tuples} i.e., rows of the multi-column table.\looseness=-1

\section{Conclusion}
We study the task of \textit{tabular query resolution}, where the user provides a partially completed table that we aim to automatically fill. 
We propose a novel framework \tabfill that directly models the latent relationship between entities in table columns through relational chains connecting them in a KB, and learns to pick the best chain amongst candidates, which is used to complete the table. 
We construct a novel and diverse data set from the WikiTable corpus~\cite{bhagavatula2015tabel} and use it to evaluate the qualitative performance of our framework, along with a detailed analysis of various error sources. 
Empirical results demonstrate the benefits of using machine learning components in respective modules.
Our proposed meta-path based entity relation representation and it's use for candidate retrieval shows significant performance improvement over traditional feature based entity retrieval strategy. 
In future, we plan to jointly lever the vast amount of online text data and the web tables, in conjunction with knowledge bases, for \textit{tabular query resolution} task to address KB incompleteness problem.
A practical deployment of \tabfill will require us to define performance metrics and implement various system optimizations, which we leave as future work.\looseness=-1

\balance
\bibliographystyle{ACM-Reference-Format}
\bibliography{paper}
%\bibliography{references}
\clearpage
\section{Appendix}

\subsection{Understanding the Data}
\label{subsection:appendixondata}

We first summarize the key properties of our custom data:
\begin{itemize}
    \item We focus on composing tables with atleast 3 rows and exactly 2 columns, where the leftmost column is the core column (i.e., all column entities are unique).
    \item All column values in each table should be linked to an external KB, in our case, Freebase.
    \item The natural language query description ($QD$) should contain the Subject Entity ($SE$) of the Table. Additionally as part of $QD$, it is desirable to have some text, describing the nature of information about $SE$.
    \item The individual length (number of edges) of $P_1$ and $P_2$ paths between any pair of entities should be atmost 3.
    \item SPARQL queries synthesized using any candidate $[P_1 - P_2]$ should satisfy pre-defined time and quality constraints to ensure practicality of our framework.
    \item A table should have at least one valid $[P_1 - P_2]$ chain satisfying all the above constraints, and can retrieve at least 2 rows of the table.  
\end{itemize}

\begin{figure}[!htb]
\includegraphics[width=\linewidth, scale=0.8]{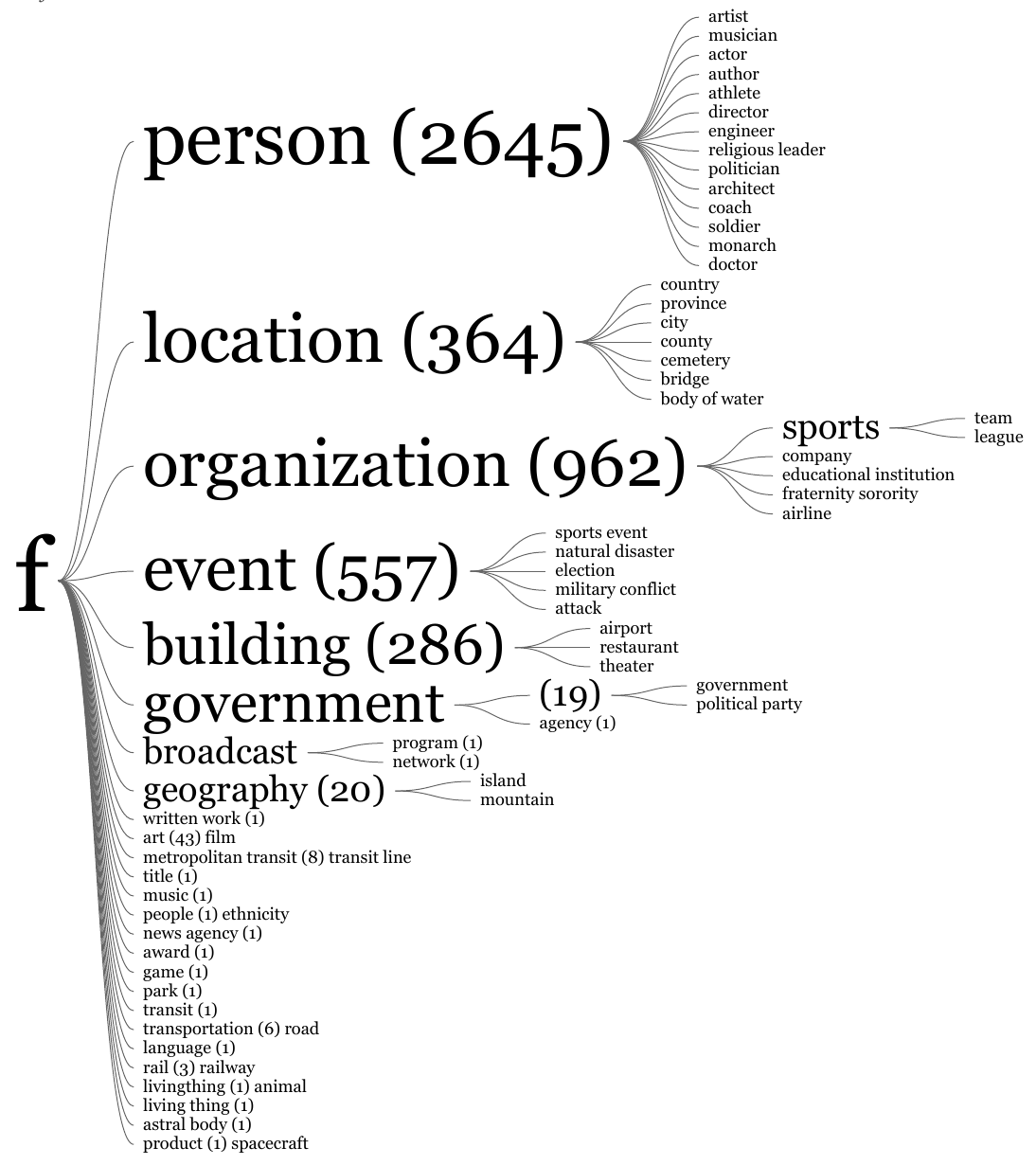}
\caption{Word Tree for the Fine Grained Entity Type of all Tables depicts the diversity of our dataset. The tree shows the hierarchy of the entity types in terms of their hypernym. Each level provides a specific (finer) type of the prior level's entity type. \looseness=-1}
\label{fig:fgetofalltables}
\end{figure}

We now present some key statistics of our dataset that satisfies all the above properties, to demonstrate that the data is \textit{diverse and challenging, while being task specific}.
We start by observing that since we use tables created in Wikipedia by human users, the tables in our data directly reflect the real world information requirement of humans.
An important aspect of our dataset is the variety of the topics of tables, as is required to create a diverse and close-to real world dataset.
While the relative distribution of SET for all 2 column tables in WikiTable corpus has not been analyzed, we focus on capturing and representing the diversity of SET in our final dataset.\looseness=-1

Figure~\ref{fig:fgetofalltables} presents a word tree\footnote{\url{ https://www.jasondavies.com/wordtree/}} of the fine grained entity type of Subject Entity of each table in our entire data.
For each fine grain entity type, we bucket each term in the FGET in terms of their hypernym. We further remove the hypernym from FGET after it has been bucketed to allow for a clean visual in understanding the distribution and hierarchy. The numbers in the parenthesis are the sum of the counts of the hyponyms that belong to that hypernym. All FGETs belong to the hypernym ``f'' and then are branched in terms of the their respective hypernym and hyponyms.
For example, the dataset contains 2645 entities that are categorised as under a person super type. The subtypes of Subject Entities that belong to the hypernym ``person'' can be artist, musician, actor, etc.
We observe that while the dominant SET includes ``person'', ``organization'' , ``event'', ``location'' etc., it also covers some relatively rare (yet important) types like ``art film'', ``government'' etc.
This bias towards specific dominant SET is intuitive, as Freebase is known to contain a lot of information on these topics, and our current dataset construction majorly focus on information contained in Freebase to select tables that can be completed using our strategy.\looseness=-1

\begin{figure}[!htb]
\includegraphics[width=\linewidth]{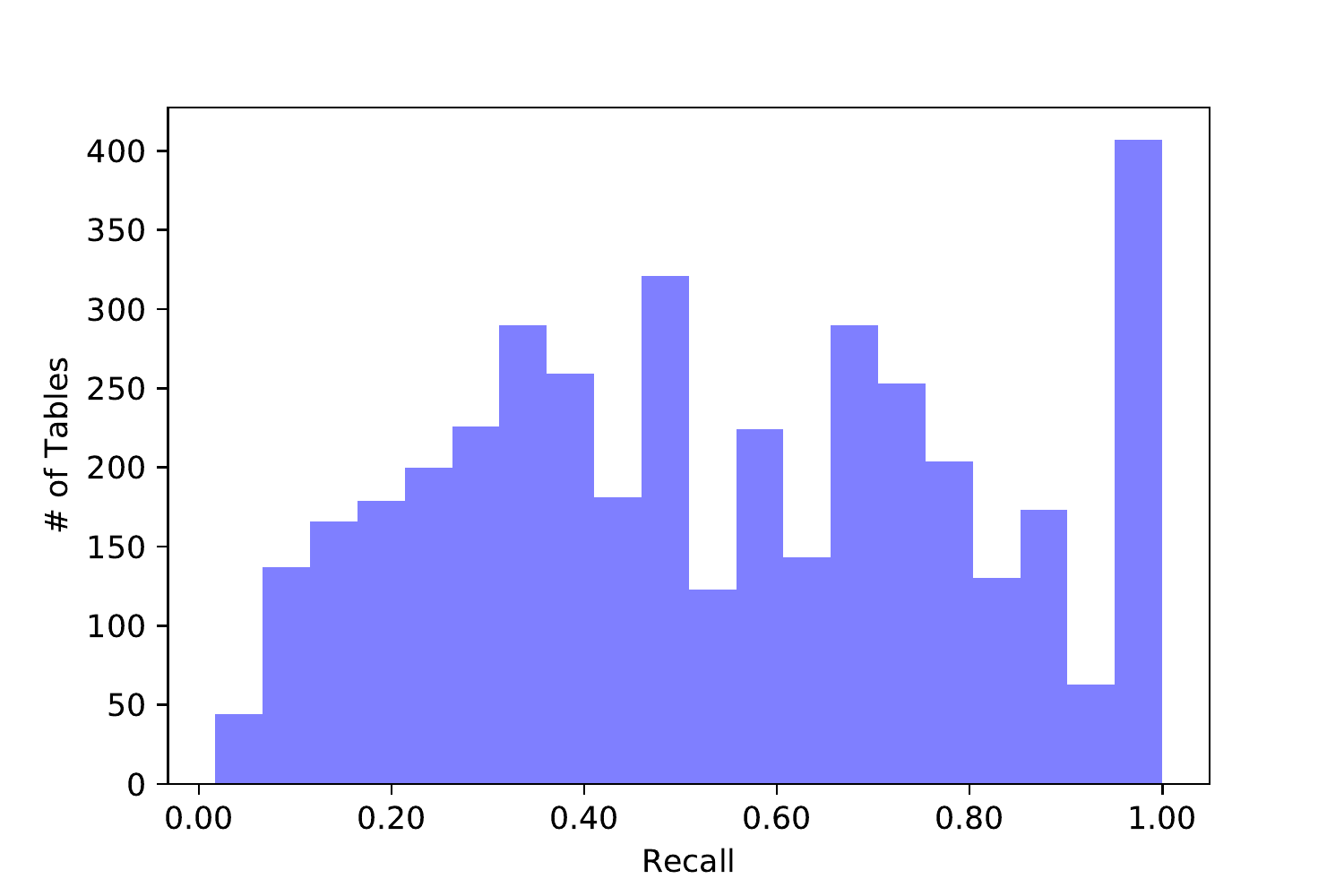}
\vspace{-20pt}
\caption{Histogram of Recall of first positive chain for each table shows that the positive chains are informative and can be used for table completion.\looseness=-1}
%\vspace{-15pt}
\label{fig:recall_hist_appendix}
\end{figure}

The primary goal of our current work is to achieve high Recall i.e., fill up as many ground truth rows as possible for any table using \tabfill.
Figure~\ref{fig:recall_hist_appendix} presents the distribution of the ground truth Recall of the first positive chain for each table in our data.
About 48.37\% of total tables have positive chain Recall $>=$ Mean Recall (0.5419), while about 9.82\% of total tables have positive chain Recall = 1.0, which means that the positive chains extracted using our strategy are meaningful for this table completion task, and that Freebase indeed has adequate information to fill-up several tables to varying degree of completeness (Recall).\looseness=-1

\begin{figure*}[!htb]
\centering
\begin{subfigure}[b]{0.48\linewidth}
\includegraphics[width=\linewidth,height=3cm]{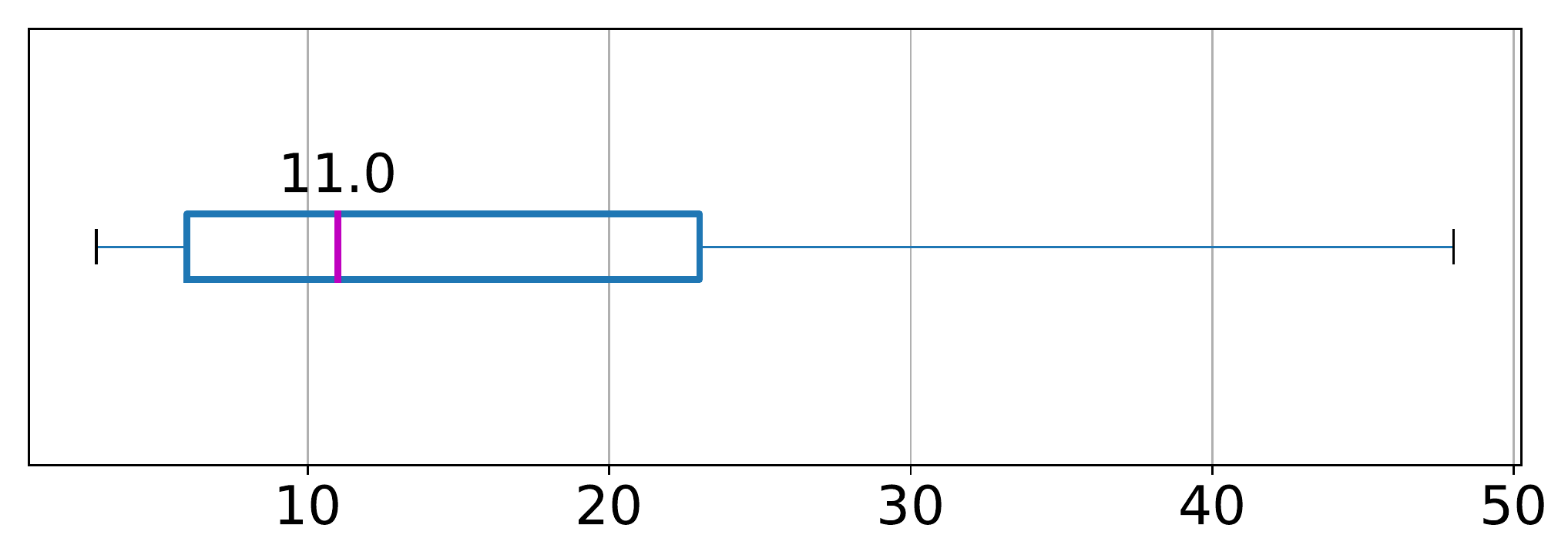}
\caption{Rows per Table without Outliers.\looseness=-1}
\label{fig:numrowsboxplotwithoutoutliers}
\end{subfigure}\qquad
\begin{subfigure}[b]{0.48\linewidth}
\includegraphics[width=\linewidth,height=3cm]{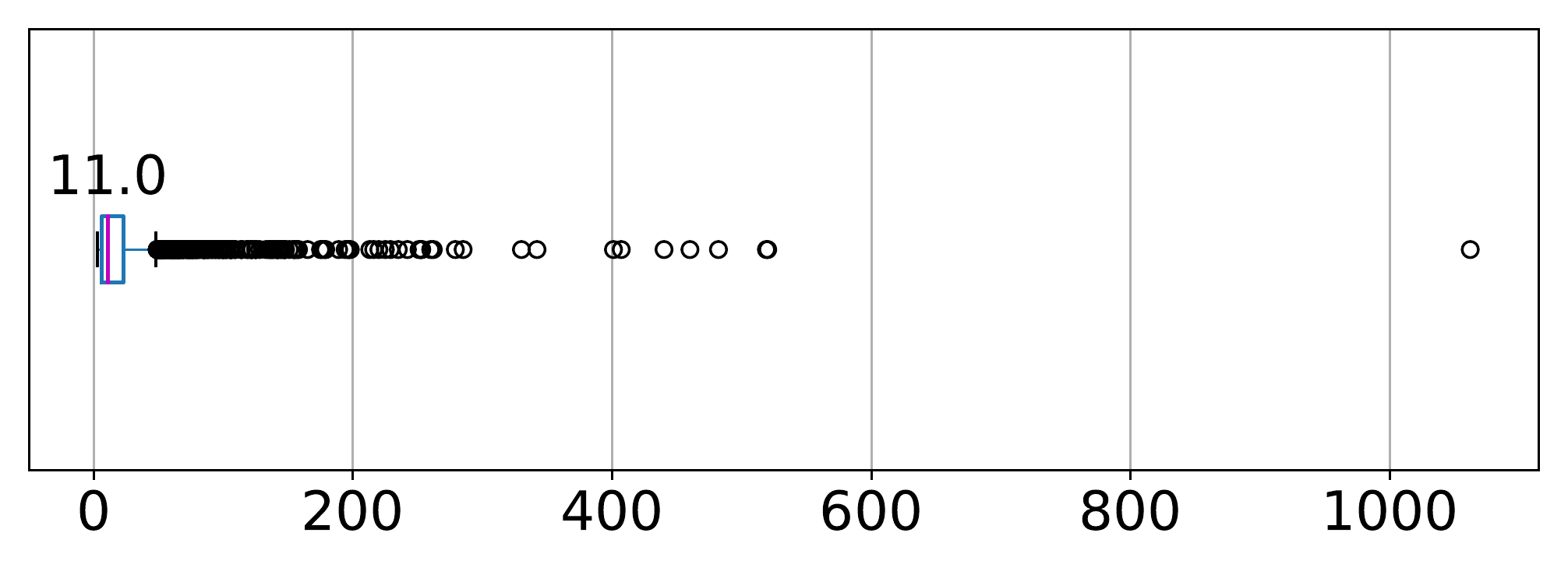}
\caption{Rows per Table with Outliers.\looseness=-1}
\label{fig:numrowsboxplotwithoutliers}
\end{subfigure}
\caption{Distribution of Number of Rows per Table with and without Outliers. The (min, 25-ile, 50-ile, mean, 75-ile, max) for the number of rows per table is $(3.0, 6.0, 11.0, 19.70, 23.0, 1062)$, with about 31.85\% tables having more rows than the mean.\looseness=-1}
\label{fig:numrowsboxplot}
\end{figure*}

Given the Recall  distribution, ideally a challenging dataset needs to have varying number of rows per table.
Figure~\ref{fig:numrowsboxplot} shows the box plot of number of rows per table in our dataset, without (Figure~\ref{fig:numrowsboxplotwithoutoutliers}) and with (Figure~\ref{fig:numrowsboxplotwithoutliers}) outliers.~\footnote{\url{https://pandas.pydata.org/pandas-docs/stable/reference/api/pandas.DataFrame.boxplot.html}}.
To summarize, all points that lie outside the range $1.5 * IQR$ are outliers (where Inter-Quartile Range IQR = 75-ile - 25-ile).
The box extends from the Q1 to Q3 quartile values of the data, with a line at the median (Q2). The whiskers extend from the edges of box to show the range of the data. The position of the whiskers is set by default to 1.5 * IQR (IQR = Q3 - Q1) from the edges of the box. Outlier points are those past the end of the whiskers.
Majority of the tables have number of rows within a compact range, as indicated by a 25-ile = 6.0, median = 11.0 and 75-ile = 23.0 respectively.
However, the mean number of rows is 19.70 (much higher than the median), with about 31.85\% tables having more rows than the mean, while max number of rows = 1062.
These tables, with skewed number of rows, are inherently hard to complete and thus form outliers in Figure~\ref{fig:numrowsboxplotwithoutliers}.\looseness=-1

\begin{figure}[!htb]
\centering
\includegraphics[width=\linewidth,height=4cm]{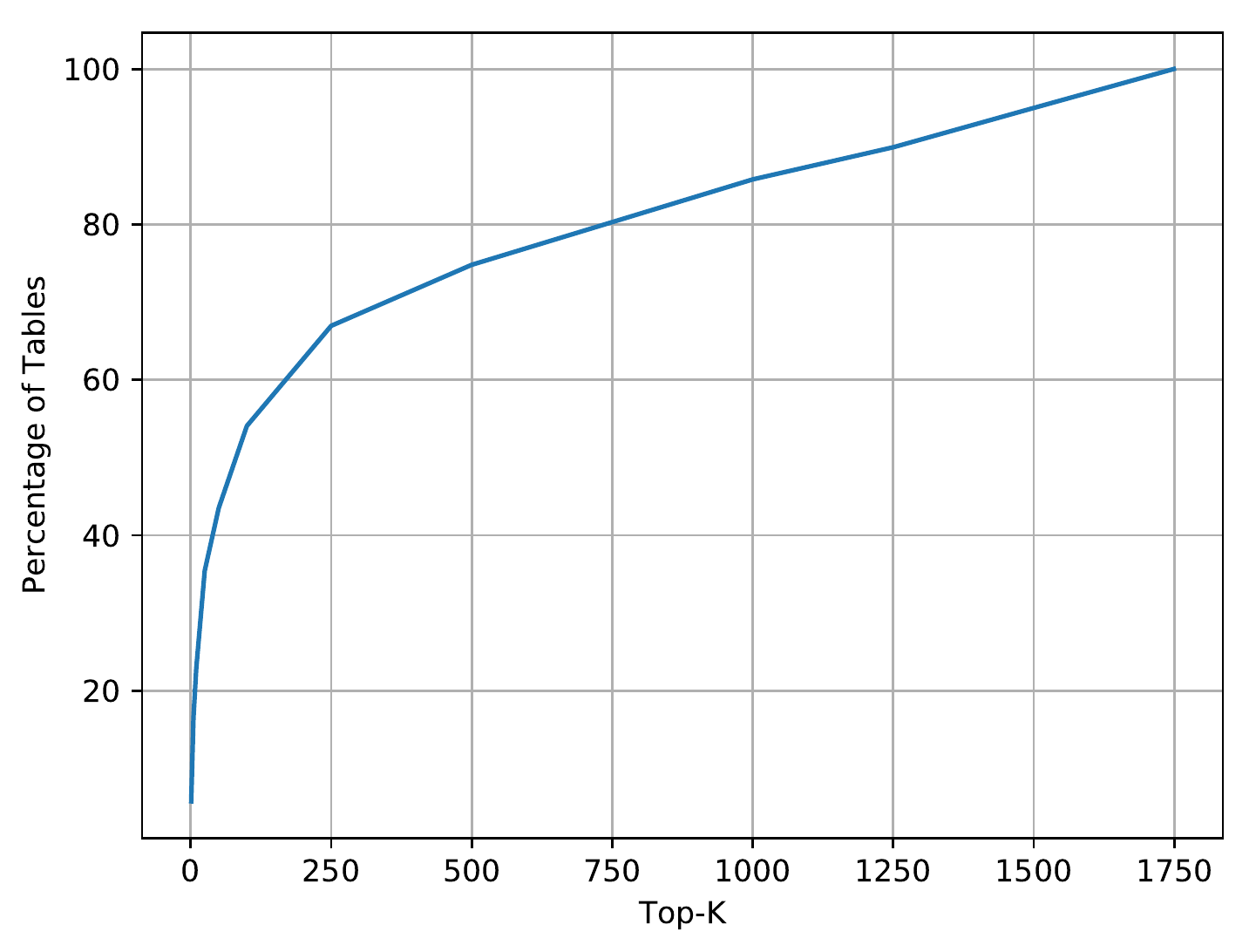}
\caption{Approximate percentage of Tables covered by Top-K most frequent positive paths. The top-10 most frequent positive paths cover $\approx$ 22.90\% tables and top-50 paths cover $\approx$ 43.53\% tables. Interestingly $\approx$ 74.80\% of the tables are covered by top-500 most frequent positive paths. \looseness=-1}
\label{fig:pospathtablecoverage}
\end{figure}

We observe that the dataset has significant skew in the distribution of the number of paths per table.
In total there are 1749 unique positive paths, 122607 unique negative paths, while 1042 paths are both positive and negative across all tables in our entire data set.
The number of paths that appear only once in the overall data is very high (e.g., 93,496 paths out of the total 123,314 (75.8\%) occur only once).
Note that this uniqueness of path is generally due to some unique sub-path (often times a single edge) or an unique combination of frequent sub-paths, which make the entire path unique.\looseness=-1

\begin{figure*}[!htb]
\centering
\begin{subfigure}[b]{0.48\linewidth}
\includegraphics[width=\linewidth,height=2.75cm]{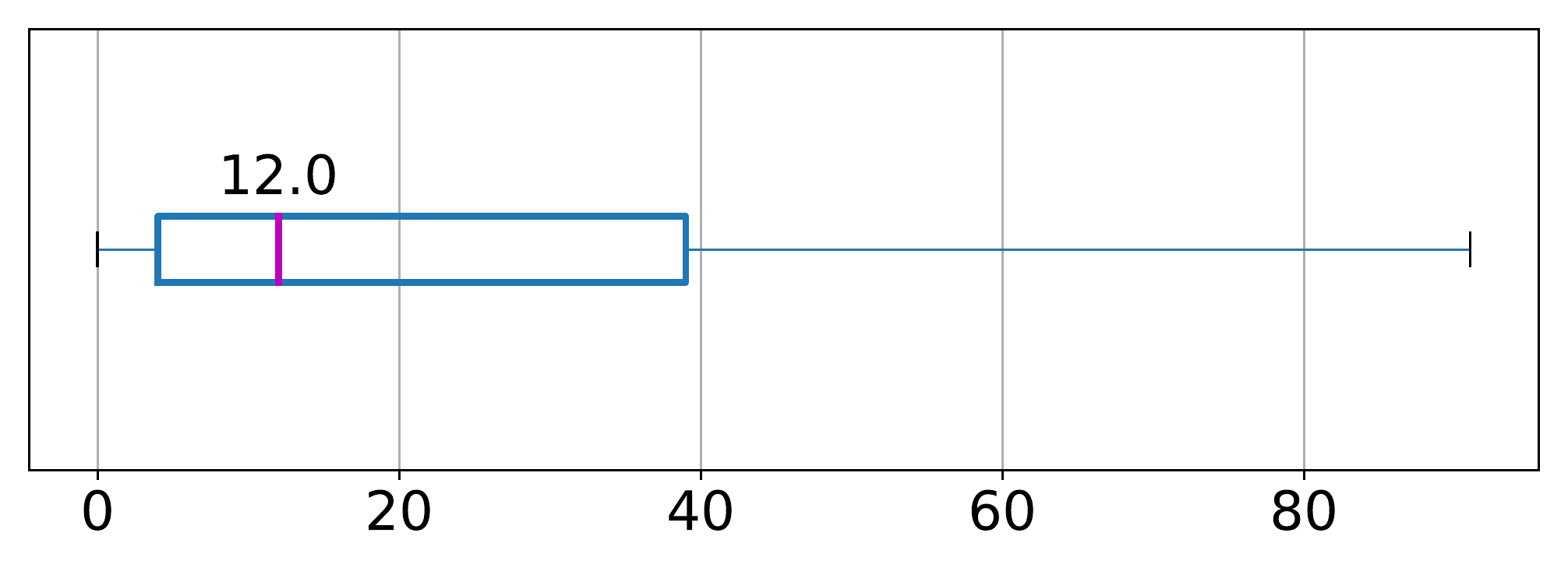}
\caption{Negative Paths per Table without Outliers.} 
\label{fig:numnegpathsboxplotwithoutoutliers}
\end{subfigure}\qquad
\begin{subfigure}[b]{0.48\linewidth}
\includegraphics[width=\linewidth,height=2.75cm]{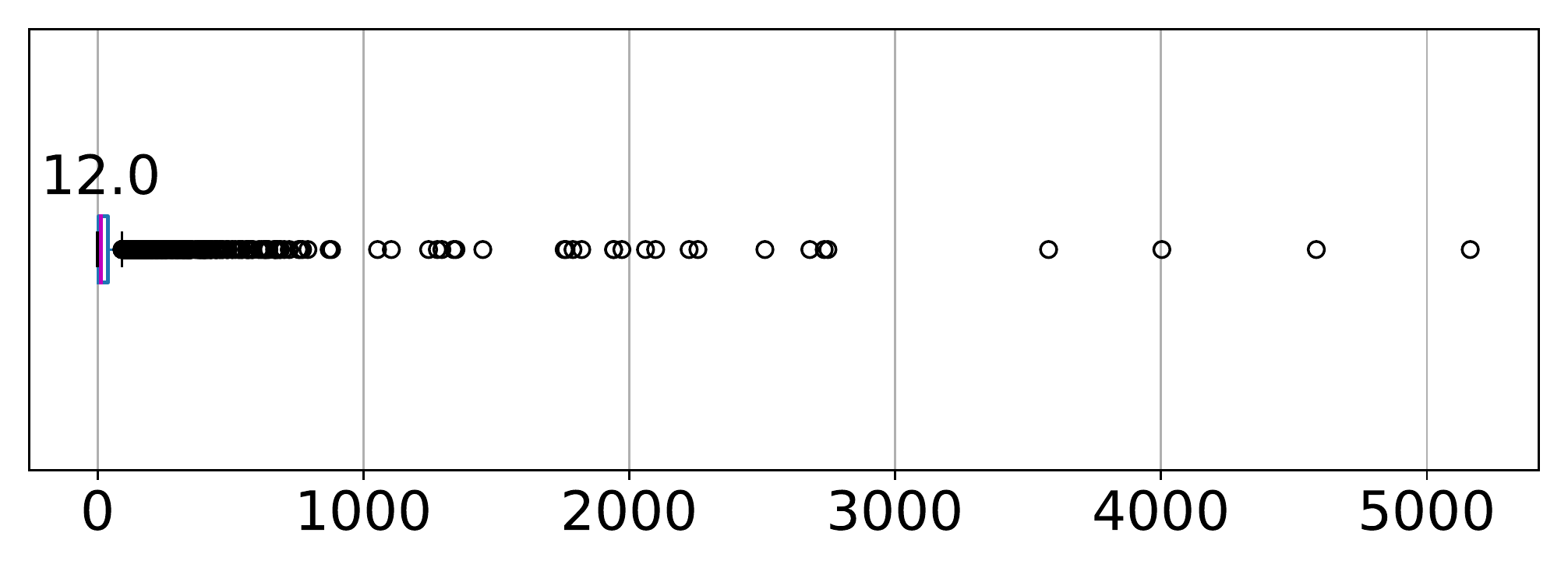}
\caption{Negative Paths per Table with Outliers.\looseness=-1}
\label{fig:numnegpathsboxplotwithoutliers}
\end{subfigure}
\caption{Distribution of Number of Negative Paths per Table with and without Outliers. The (min, 25-ile, 50-ile, mean, 75-ile, max) for the number of negative paths per table is $(0.0, 4.0, 12.0, 57.53, 39.0, 5163)$, with about 18.24\% tables having more negative paths than mean. This skew in distribution makes our data challenging, yet realistic.\looseness=-1}
\label{fig:numnegpathsboxplot}
\end{figure*}

\begin{table*}[!htb]
\centering
\small
\begin{tabular}{|p{2.9cm}|p{1.8cm}|p{4.6cm}|p{7cm}|r|}
\hline
\textbf{Query Intent String}                    & \textbf{Number(\%) of Tables} & \textbf{Sample Column Names} & \textbf{Top-1 Frequent Positive Chain}                \\ \hline
classification                   & 209 (5.21\%)             & [M] driver $\vert$ constructor    \newline [L] driver $\vert$ constructor        &  \textasciicircum{}base.formula1.formula\_1\_driver.first\_race $\sim$people.person.profession $\sim$people.profession.people\_with\_this\_profession $/$base.formula1.formula\_1\_driver.team\_member $\sim$base.formula1.formula\_1\_team\_member.team \\ \hline

airlines destinations            & 136 (3.39\%)             & [M] airline $\vert$ destination   \newline [L] airline $\vert$ destination        &  aviation.airport.hub\_for $\sim$business.business\_operation.industry $\sim$business.industry.companies $/$aviation.airline.hubs  \\ \hline

discography singles              & 115 (2.87\%)             & [M] title title $\vert$ album album   \newline [L] single detail single detail single detail $\vert$ album album album &    music.artist.track$\sim$music.recording.song$/$ music.composition.recordings$\sim$music.recording.releases $\sim$\textasciicircum{}music.album.primary\_release  \\ \hline

race                             & 95 (2.37\%)              & [M] driver $\vert$ constructor    \newline [L] driver $\vert$ constructor         &    \textasciicircum{}base.formula1.formula\_1\_driver.first\_race $\sim$people.person.profession $\sim$people.profession.people\_with\_this\_profession $/$base.formula1.formula\_1\_driver.team\_member $\sim$base.formula1.formula\_1\_team\_member.team  \\ \hline

qualifying                       & 92 (2.29\%)              & [M] driver $\vert$ constructor    \newline [L] driver $\vert$ team               &     \textasciicircum{}base.formula1.formula\_1\_driver.last\_race $\sim$people.person.profession $\sim$people.profession.people\_with\_this\_profession $/$base.formula1.formula\_1\_driver.team\_member $\sim$base.formula1.formula\_1\_team\_member.team \\ \hline

filmography                      & 74 (1.84\%)              & [M] film $\vert$ language     \newline [L] television title $\vert$ television note	                 & film.actor.film $\sim$film.performance.film$/$film.film.language        \\ \hline

television                       & 67 (1.67\%)              & [M] title $\vert$ network     \newline [L] drama drama $\vert$ drama broadcasting network            & people.person.nationality$\sim$\textasciicircum{}tv.tv\_program.country\_of\_origin $/$tv.tv\_program.original\_network$\sim$tv.tv\_network\_duration.network  \\ \hline

singles                          & 56 (1.4\%)               & [M] title title $\vert$ album album  \newline [L] title title $\vert$ album ep    album ep        &    music.artist.track$\sim$music.recording.song $/$music.composition.recordings$\sim$music.recording.releases $\sim$\textasciicircum{}music.album.primary\_release  \\ \hline

numtkn season opening day lineup & 48 (1.2\%)               & [M] opening day starter name $\vert$ opening day starter position \newline [L] player $\vert$ position     &   \textasciicircum{}baseball.batting\_statistics.team $\sim$baseball.batting\_statistics.player $/$baseball.baseball\_player.position\_s  \\ \hline

passenger                        & 48 (1.2\%)               & [M] airline $\vert$ destination \newline [L] airline $\vert$ destination                                   &  aviation.airport.hub\_for $\sim$business.business\_operation.industry $\sim$business.industry.companies $/$aviation.airline.hubs \\\hline             
\end{tabular}
\caption{List of Top-10 most frequent intent string and their approximate table coverage. We sample the most and least frequent column names as well as the top-1 most frequent positive chain for set of tables containing those intent string. For each intent string, we first gather the number of tables that correspond to that intent string and percentage of the tables out of the whole dataset. The aggregated column names of each table in this subset (corresponding to that intent string) are shown. The [M] indicates the most frequent column names and the [L] indicates the least frequent column names. If these column names are the same, it means all columns in the subset of tables have the same name. The top$-$1 frequent positive chain for each such table sub-groups are presented in the last column. The "/" is used to separate the $P_1$ sub-path from the $P_2$ sub-path of the $[P_1 - P_2]$ chain (for better readability). \looseness=-1  }
\label{tab:top10intentstring}
\end{table*}

The \textit{median number of positive paths} per table is 1.0 while the \textit{mean} is 1.26, with \textit{maximum} being 90, which makes the distribution of positive paths per table less skewed.
However, the skew in the number of negative paths per table (and hence total candidate paths per table) is significant as is observed in Figure~\ref{fig:numnegpathsboxplotwithoutliers} with outliers,  compared to Figure~\ref{fig:numnegpathsboxplotwithoutoutliers} without outliers.
To summarize, the \textit{(min, 25-ile, 50-ile, mean, 75-ile, max)} for the number of negative paths per table is (0.0, 4.0, 12.0, 57.53, 39.0, 5163), where about 18.24\% tables having more negative paths than mean.
The skew  in the number of negative paths, as well as the overlap of labels for several paths,  make the learning task for the QTS module particularly difficult.
To add to this challenge, we have a skew in the percentage of tables covered by frequent positive paths  as well.
As observed in Figure~\ref{fig:pospathtablecoverage}, the top-500 most frequent positive paths cover about 74.80\% of the tables, which means, some of the remaining tables have positive paths that are infrequent in the data.\looseness=-1

The Wikipedia pages are created by humans and hence there is a tendency to create ``template'' pages.
Thus for tables with similar information for related types of entities, we expect similar Page Title, Table Caption, Column names in the Wikipedia pages. For example, we observe that there are several pages with the template title ``Artist X discography'', template table caption ``Singles'', and column names as ``(Title Title, Album Album)'', for many different singer ``X''. A detailed list of top-10 most frequent Query Intent String along with sample most [M] and least [L] frequent column names, and top-1 most frequent positive paths is presented in Table~\ref{tab:top10intentstring}. This list is generated by first aggregating all tables in the dataset by their QIS. For each of the top 10 QIS, we count the number of tables that have a matching intent  string. This count is then used to find the percentage of tables (out of the whole dataset) that belongs to each QIS. For a table that belongs to a particular QIS, we collect its column names and  positive path chains. Next, we sort the column names and positive path chains (collected for each table with a matching QIS) by their frequency to report the most and least frequent occurrence. Note, the most and least frequent column names can be the same, thereby indicating no variance in the column names for such tables.\looseness=-1

\begin{figure*}[!htb]
\centering
\begin{subfigure}[t]{0.78\linewidth}
\begin{center}
% trim=10.5cm 10.5cm 10.5cm 10.5cm
\includegraphics[width=\linewidth,height=0.32\textheight,scale=1.0]{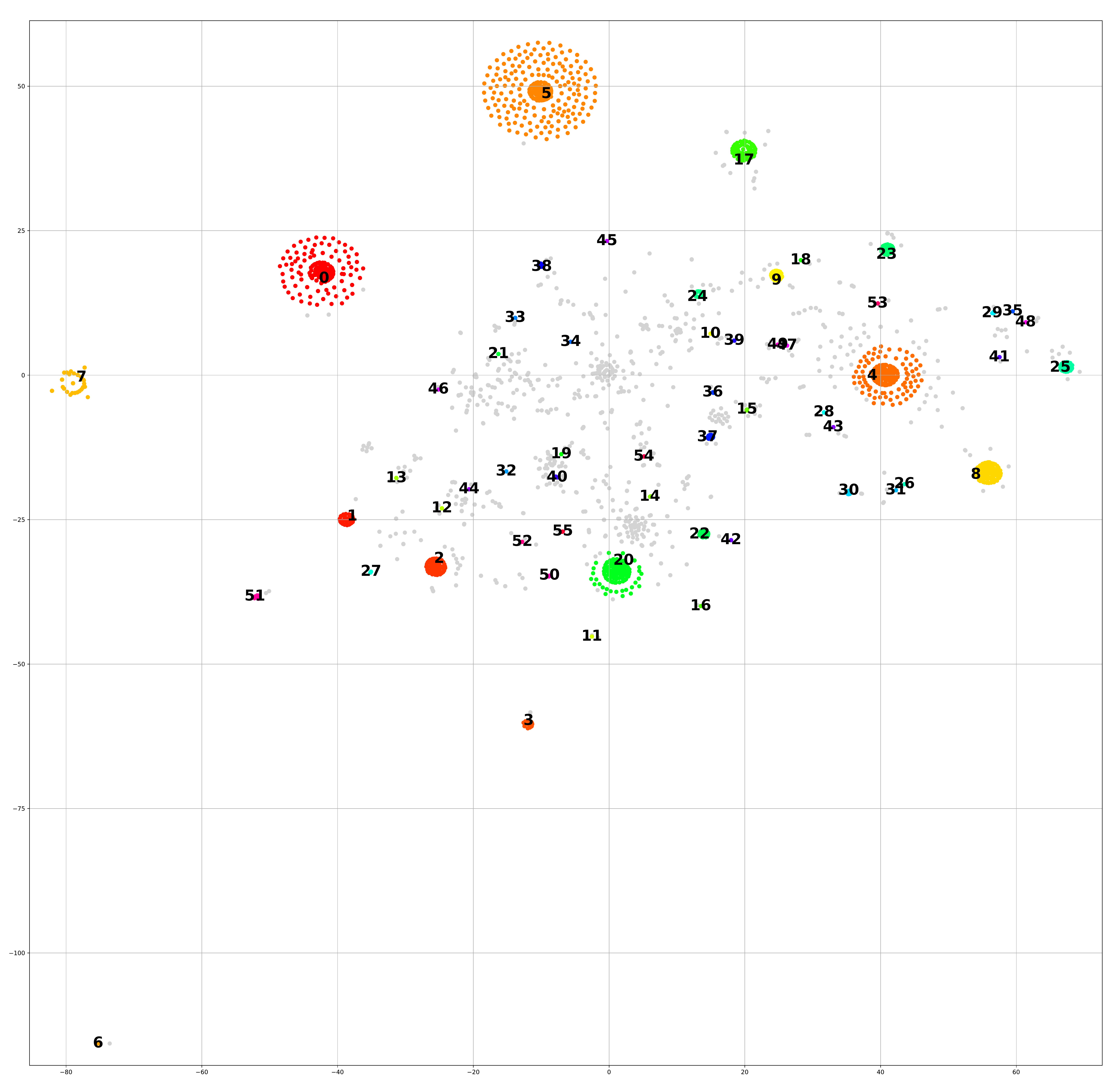}
\end{center}
\label{fig:clusters}
\end{subfigure}
\begin{subfigure}{\textwidth}
\caption{DBScan clustering (with similarity $=$ cosine, min pts $=$ 10 , eps $=$ 0.28) followed by t-SNE (with perplexity $=$ 40,  n\_components $=$ 2, init $=$ pca, n\_iter $=$ 2500, random\_state $=$ 23)
visualization of the clusters created using the BOW representation (Count Vectors) of tokens in only the column names. We obtain 56 clusters covering $\approx$ 66\% of the tables, while the remaining $\approx$ 34\% tables end up as \textit{Noise points} after clustering. The \textbf{dominant} i.e., most frequent tokens of each cluster is presented in the below Legend diagram.\looseness=-1}
\end{subfigure}
\begin{subfigure}{0.5\textwidth}
  \centering
  \includegraphics[width=.65\linewidth, height=0.55\textheight]{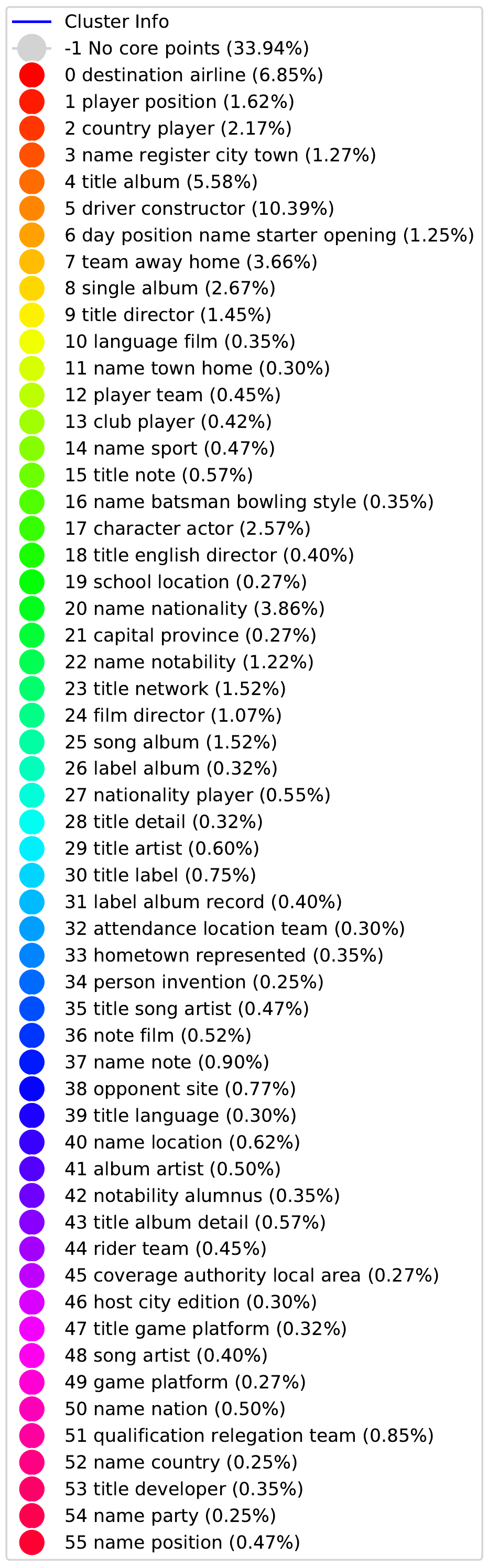}
  \label{fig:clusterinfo}
\end{subfigure}%
\begin{subfigure}{.5\textwidth}
  \centering
  \includegraphics[width=.65\linewidth, height=0.55\textheight]{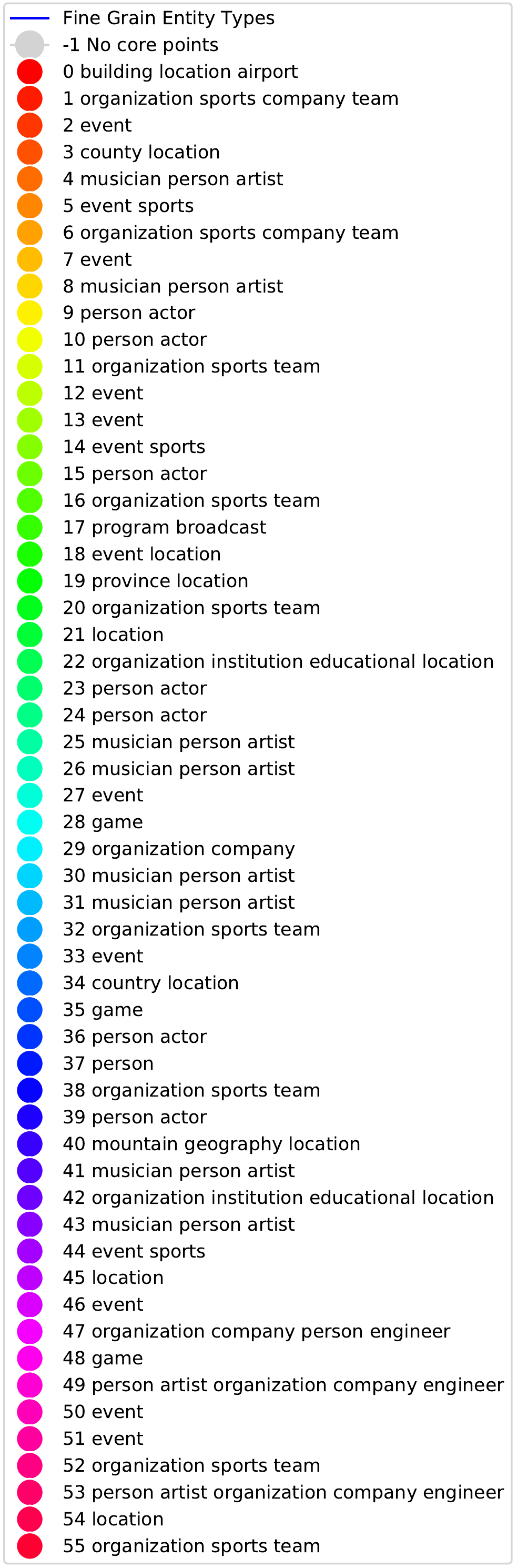}
  \label{fig:clusterfget}
\end{subfigure}
\begin{subfigure}{\textwidth}
  \caption{Each cluster id is labeled using the dominant tokens in column name field for the set of Core points in that cluster, under the \textit{Cluster info} column on the left. The number in the bracket denotes an approximate percentage of number of tables in each cluster wrt whole data. The right column contains the dominant tokens of the $FGET$ of the tables, which constitute core points for the corresponding cluster id.\looseness=-1}
\end{subfigure}
\caption{Visualization of data clustering using column names and corresponding FGET of clusters.}
\label{fig:dataclustering}
\end{figure*}

To further understand the extent of learnable information in our data, and to assess the impact of noise due to natural language variations to express the same intent, we perform clustering of column names in our dataset, and label the clusters with the column names and Fine Grained Entity Type (FGET) of the core point of each cluster, shown in Figure~\ref{fig:dataclustering}.
For each table in our dataset, we extract the column name pairs and treat them as a single string.
We also extract the FGET for each table as a string with unique tokens (i.e. `` person actor person'' is treated as ``person actor'') and remove type ``f'' from the string (as this type is a part of every FGET).
Each column name string for each table is treated as a document and vectorized using scikit-learn's CountVectorizer (with binary $=$ true) \footnote{\url{https://scikit-learn.org/stable/modules/generated/sklearn.feature_extraction.text.CountVectorizer.html}}.
We apply DBScan clustering to the vectorized documents and use the core points of each cluster to label the clusters. 
For each core point of a cluster, we curate a list of column names and a list of FGET. 
The highest frequency column name and FGET are used as the cluster label in each respective legend.
The text in the legends is truncated to 50 characters to increase readability. 
We label all noise points (cluster -1) as ``No core point''.
The final clustering is shown in Figure~\ref{fig:dataclustering}a, and the corresponding most frequent column name for each cluster as well as the FGET of the corresponding cluster is shown in Figure~\ref{fig:dataclustering}b.\looseness=-1

There are several clusters being formed on the related set of tables (with similar FGET) using the natural language part of inputs i.e., Column Names (CN) in this case.
While some of those are clean clusters, other have a lot of noise points around them.
There are a few strong clusters (eg: \#3 and \#7, which jointly covers $\approx$ 17\% tables) with hardly any noise points around them, indicating that tables with these FGET have almost identical column names, resulting in neat clustering.
For example, cluster \#8 has column names ``album, singles'' with the FGET as ``person musician artist'', and covers $\approx$2.67\% of the tables.
This cluster has a few Noise points detected by the DBScan algorithm  (although much less than many other clusters), which is caused due to the variation in the natural language description of column names for tables with FGET as ``person musician artist''. 
Note that there is another dedicated cluster (\# 25) with the exact same FGET as cluster\# 8, representing similar topic for the table, but with totally different column names i.e., ``label, title'' in this case.
Natural language variations to represent similar information is evident in our dataset from some of the other clusters as well.
Also, $\approx$ 34\% of the tables in the data are flagged as noise points (i.e., without any core points) by DBScan algorithm, indicating relatively high amount of natural language variations present in our data.
Note that of the total $\approx$ 66\% tables with cluster-able column names (i.e., relatively non-noisy column names), we observe a skewed distribution of column names across the clusters, as indicated by the variation in the cluster size i.e., percentage of tables within each cluster.
We noticed similar trends while analyzing clusters obtained by combining various information sources in our data (eg: query intent string, column names etc.), which resulted in higher number of smaller sized clusters with increased variability in the natural language cluster descriptors.
These diversity make our data a good representative of real world scenario, while also making it challenging to work with.

\subsection{QTS Module Errors}
\label{subsection:qtsmoduledebugging}

Here, we present additional details of the failures of individual components of QTS to select positive chain for a table, and briefly mention about our future work to improve this component.
The JacSim scorer, while being naive compared to learning modules, gives an estimate of how many tables in the Validation/Testing set are easy enough, for which the positive chain can be predicted correctly using a non-learning based component. 
We first analyze its failures, and then focus on analyzing the common set of failing tables for the learning based modules.
Note that, a common issue for all our score based matching technique is that, given a $<$QD, CN, SET$>$ and a set of candidate chains $CC$, sometimes the model assigns same score to several chains in \textit{ CC} (eg: positive chain as well as the first few negative chains for that table), in which case the performance degrades to random behaviour.\looseness=-1

\textbf{\noindent Analyzing failures of JacSim :} Key reasons are:
\begin{itemize}
    \item The SET have more matches with the candidate chains as they both use KB\_Vocab. Thus the $<$QD, CN$>$  part has lesser importance in the overall computed score.
    \item Jaccard similarity focuses more on common (matching) tokens, which can be very generic and noisy, and thus fails to focus on the special tokens (which often have valuable information).
    \item JacSim completely fails when the positive path has few or no matching tokens with $<$QD, CN, SET$>$ compared to the negative ones, which can have too generic yet more matching tokens (thus false positives).
    \item A rare scenario occurs when negative chains have same overlapping tokens as the positive chain, but the positive chain has more total tokens overall (denominator is higher) causing its JacSim score to be lower.\looseness=-1
\end{itemize}

\textbf{\noindent Analyzing common failures for LR, KNN, RF, DNN:}
We observe in Table~\ref{tab:queryselection} that the performance of all the 4 models somewhat thresholds, with a maximum Accuracy@1 of 57.67 in validation and 55.94 in testing data for the DNN model, but higher mean Table\_Recall for the RF model compared to the Oracle.
We observe that there is a significant overlap in the set of tables in respective partitions for which all these models consistently fail to predict the best positive chain even across 10 simulations.
Hence, we further analyze the tables in the common failure set ($\approx$ 33\% for both Validation and Testing) to understand the key reasons for failure.\looseness=-1

\begin{table}[!htb]
\small
%\centering
\begin{tabular}{|p{4cm}|p{1.75cm}|p{1.75cm}|}
\cline{1-3}
\textbf{Data Characteristic} & Validation & Testing\\
\cline{1-3}
\hline
Total Tables Tested & 378 & 379\\
\hline
Total Common Failing Tables & 123 (32.54\%) & 125 (32.98\%)\\
\hline
Tables with OOV token in QD & 20 (16.26\%) & 34 (27.2\%)\\
\hline
Tables with OOV token in CN & 5 (4.07\%) & 21 (16.8\%)\\
\hline
Tables with OOV token in $<$QD, CN$>$ & 23 (18.70\%) & 41 (32.8\%)\\
\hline
Tables with OOV token in SET & 2 (1.63\%) & 11 (8.8\%)\\
\hline
Total Unique Positive $P_1$ & 100 & 111\\
\hline
Total Unique Positive $P_2$ & 81 & 93\\
\hline
Total Unique Positive $P_1 - P_2$ & 121 & 132\\
\hline
Num Positive Paths per Table \newline [50-ile, Mean, Max] & [1,1.13,3] & [1,1.22,4]\\
\hline
Num Negative Paths per Table \newline [50-ile, Mean, Max] & [31,126.33,2510] & [36,104.3,2747]\\
\hline
Tables with infrequent Positive $P_1 - P_2$ & 27 (21.95\%) & 34 (27.2\%)\\
\hline
Tables with Positive $P_1 - P_2$ that have dual labels in the Training data set & 93 (75.61\%) & 90 (72.0\%)\\
\hline
Tables with infrequent Positive $P_1 - P_2$ and OOV token in SET & 1 (0.8\%) & 1 (0.8\%)\\
\hline
Tables with infrequent Positive $P_1 - P_2$ and rare token in $<$QD, CN$>$ & 12 (9.76\%) & 20 (16.0\%)\\
\hline
\end{tabular}
\caption{Statistics of common set of Tables on which KNN, LR, RF, DNN models failed consistently in 10 simulations.\looseness=-1}
\label{tab:qtserroranalysis}
\end{table}

Table~\ref{tab:qtserroranalysis} presents some of the key data statistics for the common failing set of tables.
We observe that several tables in Validation and Testing data contain the Out of Vocabulary (OOV) token in various information fields i.e., natural language and knowledge base respectively.
The OOV tokens are replacement for rare tokens that appear only once in the Training and Validation data combined in the respective fields, and hence has been removed from the corresponding vocabulary (Table\_Vocab and KB\_Vocab respectively).
We observe that the number of tables with OOV tokens in $<$QD, CN$>$ in Testing data is much more than that of Validation data, and quite high overall ($\approx$ 33\%).
This is caused by the sparsity of the natural language tokens in the $QD$ and $CN$ fields, with $QD$ having more rare tokens than $CN$.
On the contrary, the SET, as well as the positive chains, seem to be much less impacted by the OOV, when compared to the natural language fields.
It is somewhat intuitive as a natural language token can have several possible variations of itself, some of which might be rare in the corpus hence removed, while the KB token set is generally fixed due to the static nature of KB and KB tokens have a tendency to co-occur together as part of a type or sub-path.
Note, sometimes the rare tokens (which are replaced by the OOV token) are more informative than the common tokens for decision making purpose by QTS, as the rare tokens contain the \textit{special granular information} that best describes the relationships in the current table.\looseness=-1

Next, we notice that the mean number of negative paths per table in the common failure set is 126.33 in Validation and 104.3 in Testing set, which is much higher than the mean number of total paths (positive and negative combined) for all tables in the respective partitions (68.03 in validation and 59.53 in testing).
Thus this common failure set has a significant skew in distribution of negative paths.
To add to the problem, there are 72\% of tables in Testing set which have atleast one positive path that has appeared as both positive and negative path in the training data.
In some cases, the $QD$, $CN$ and sometimes even the SET of such tables are same, but due to KB incompleteness problem, the same path has different labels in our corpus, which confuses the models.
Importantly, it is the $P_1$ part of the $P_1 - P_2$ path, which is seen to be more diverse, as indicated by the number of unique $P_1$, $P_2$ and $P_1 - P_2$ paths in Table~\ref{tab:qtserroranalysis}.
In future we plan to include some additional features as input to the QTS module, for each of the Subject Entity as well as the entities in the Example Row, by leveraging the topological and contextual information of those entities from Freebase.
These additional features can be used in conjunction with some Attention mechanism built into our DNN model, which can effectively select the $P_1$ part of the path, which is often more unique (noisy) than $P_2$ due to KB incompleteness.\looseness=-1

Another important source of error is the infrequent $P_1 - P_2$ positive path, which is present for $\approx$ 27\% of tables in Testing set.
To put in perspective, our dataset has a very high number of $P_1 - P_2$ paths that are infrequent i.e., appear only once in the overall data (e.g., 93,496 paths out of the total 123,314 (75.8\%) occur only once).
The paths themselves are composed of relationship sub-paths : $P_1 = R_1^1/R_2^1/../R_m^1$ and $P_2 = R_1^2/R_2^2/../R_n^2$ in KB, where $m$ and $n$ are lengths of the individual paths and $R_i$ is the relationship label text.
Thus an unique token in $R_i$ differentiates it from other relationships with common prefix, while an unique combination of $R_i$'s in a specific sequence make $P_1 - P_2$ unique as a whole.
Firstly{\color{blue},} our DNN (or other) model does not explicitly capture the sequential nature of relations, for which we plan to use LSTM style encoders for ETE and PCE in future.
Secondly, the individual $R_i$'s need to be encoded first as a sequence and then the individual sequence encodings need to be combined using a hierarchical architecture, which can better encode the sequential relation and partially address the path uniqueness (sparsity) problem, which we leave as future work.\looseness=-1

\subsection{Discussion on Performance}
\label{subsection:discussions}

The training of the machine learning modules is the most time consuming operation, but typically is a one-time task and can be done offline.
Currently we have assumed that all data pre-processing tasks (i.e., tasks of Query Processor module) has already been done offline during data construction.
In real world scenario given a tabular query, while there will be some additional time needed for the entity linking and text processing tasks, we anticipate the main performance bottleneck to be in Freebase operations.
Such operations include a) path search (within 3-hops) for input tuple pairs by QP, b) execution of synthesized SPARQL query by QTS, and c) for retrieving entity specific information (like description, rdf types etc.) by CTR.
All these Freebase tasks require a scaled out deployment of Virtuoso server in a cluster with high bandwidth and fast I/O, coupled with careful tuning of the server configurations (currently assumed constant), which we leave as future work.\looseness=-1

The other time consuming operations are in the CTR module, whose performance is dependent on two factors: 1) the total number of candidate rows retrieved by the SPARQL query selected by the QTS module and 2) the individual entity's information in Freebase (e.g. non-empty description, number of types etc.).
Our framework is tuned to select high Recall meta-paths so that we do not miss out relevant rows and hence some of the paths can be of low Precision.
This results in high variability in number of candidate rows for the ranker thereby skewing the time taken by it.
Currently, it takes approximately 0.02 secs to construct all the features for one row (i.e., entity pair) by the CTR, given that we pre-load all relevant entity information in memory.
It can be further reduced by removing some low importance features (discussed in Section~\ref{subsection:ctrerror}), while also leveraging both caching as well as thread level parallelism during feature construction, which we leave as future work.\looseness=-1
\end{document}